%% file: gaugestructure.tex
\input ruled
\input youngtab
\input epsf

\newfam\scrfam
\batchmode\font\tenscr=rsfs10 \errorstopmode
\ifx\tenscr\nullfont
        \message{rsfs script font not available. Replacing with calligraphic.}
        \def\scr{\cal}
\else   
        \font\sevenscr=rsfs7
        \font\fivescr=rsfs5
        \skewchar\tenscr='177 \skewchar\sevenscr='177 \skewchar\fivescr='177
        \textfont\scrfam=\tenscr \scriptfont\scrfam=\sevenscr
        \scriptscriptfont\scrfam=\fivescr
        \def\scr{\fam\scrfam}
        \def\cal{\scr}
\fi
\catcode`\@=11
\newfam\frakfam
\batchmode\font\tenfrak=eufm10 \errorstopmode
\ifx\tenfrak\nullfont
        \message{eufm font not available. Replacing with italic.}
        \def\frak{\it}
\else
	
	\font\sevenfrak=eufm7 \font\fivefrak=eufm5
	\textfont\frakfam=\tenfrak
	\scriptfont\frakfam=\sevenfrak \scriptscriptfont\frakfam=\fivefrak
	\def\frak{\fam\frakfam}
\fi
\catcode`\@=\active
\newfam\msbfam
\batchmode\font\twelvemsb=msbm10 scaled\magstep1 \errorstopmode
\ifx\twelvemsb\nullfont\def\Bbb{\bf}
        
	\font\eightbbb=cmb10 at 8pt
	\message{Blackboard bold not available. Replacing with boldface.}
\else   \catcode`\@=11
        \font\tenmsb=msbm10 \font\sevenmsb=msbm7 \font\fivemsb=msbm5
        \textfont\msbfam=\tenmsb
        \scriptfont\msbfam=\sevenmsb \scriptscriptfont\msbfam=\fivemsb
        \def\Bbb{\relax\expandafter\Bbb@}
        \def\Bbb@#1{{\Bbb@@{#1}}}
        \def\Bbb@@#1{\fam\msbfam\relax#1}
        \catcode`\@=\active
	
	\font\eightbbb=msbm8
\fi
        \font\fivemi=cmmi5
        \font\sixmi=cmmi6
        \font\eightrm=cmr8              \def\xrm{\eightrm}
        \font\eightbf=cmbx8             \def\xbf{\eightbf}
        \font\eightit=cmti10 at 8pt     \def\xit{\eightit}
                
        \font\eighttt=cmtt8             \def\xtt{\eighttt}
        \font\eightcp=cmcsc8
        \font\eighti=cmmi8              \def\xold{\eighti}
        \font\eightmi=cmmi8
        \font\eightib=cmmib8             \def\xbold{\eightib}
        \font\teni=cmmi10               \def\old{\teni}
        \font\tencp=cmcsc10

        \font\twelvecp=cmcsc10 scaled\magstep1
        
        \font\sixrm=cmr6
        \font\fiverm=cmr5

        \font\eightsy=cmsy8
        \font\sixsy=cmsy6
        \font\eightsl=cmsl8
        \font\sixbf=cmbx6

	 at10pt	
	\font\twelvehelvbold=phvb at12pt
	 at14pt
	\font\sixteenhelvbold=phvb at16pt

\def\noblackbox{\overfullrule=0pt}
\noblackbox

\def\eightpoint{
\def\rm{\fam0\eightrm}
\textfont0=\eightrm \scriptfont0=\sixrm \scriptscriptfont0=\fiverm
\textfont1=\eightmi  \scriptfont1=\sixmi  \scriptscriptfont1=\fivemi
\textfont2=\eightsy \scriptfont2=\sixsy \scriptscriptfont2=\fivesy
\textfont3=\tenex   \scriptfont3=\tenex \scriptscriptfont3=\tenex
\textfont\itfam=\eightit \def\it{\fam\itfam\eightit}
\textfont\slfam=\eightsl \def\sl{\fam\slfam\eightsl}
\textfont\ttfam=\eighttt \def\tt{\fam\ttfam\eighttt}
\textfont\bffam=\eightbf \scriptfont\bffam=\sixbf 
                         \scriptscriptfont\bffam=\fivebf
                         \def\bf{\fam\bffam\eightbf}
\normalbaselineskip=10pt}

\newtoks\headtext
\headline={\ifnum\pageno=1\hfill\else
	\ifodd\pageno{\eightcp\the\headtext}{ }\dotfill{ }{\old\folio}
	\else{\old\folio}{ }\dotfill{ }{\eightcp\the\headtext}\fi
	\fi}
\def\makeheadline{\vbox to 0pt{\vss\noindent\the\headline\break
\hbox to\hsize{\hfill}}
        \vskip2\baselineskip}
\newcount\infootnote
\infootnote=0
\def\foot#1#2{\infootnote=1
\footnote{${}^{#1}$}{\vtop{\baselineskip=.75\baselineskip
\advance\hsize by
-\parindent{\eightpoint\rm\hskip-\parindent #2}\hfill\vskip\parskip}}\infootnote=0}
\newcount\refcount
\refcount=1
\newwrite\refwrite
\def\oldsize{\ifnum\infootnote=1\xold\else\old\fi}
\def\ref#1#2{
	\def#1{{{\oldsize\the\refcount}}\ifnum\the\refcount=1\immediate\openout\refwrite=\jobname.refs\fi\immediate\write\refwrite{\item{[{\xold\the\refcount}]} 
	#2\hfill\par\vskip-2pt}\xdef#1{{\noexpand\oldsize\the\refcount}}\global\advance\refcount by 1}
	}
\def\refout{\eightpoint\catcode`\@=11
        \xrm\immediate\closeout\refwrite
        \vskip2\baselineskip
        {\noindent\twelvecp References}\hfill\vskip\baselineskip
        \baselineskip=.75\baselineskip
        \input\jobname.refs
        \baselineskip=4\baselineskip \divide\baselineskip by 3
        \catcode`\@=\active\rm}

\def\skipref#1{\hbox to15pt{\phantom{#1}\hfill}\hskip-15pt}

\def\hepth#1{\href{http://xxx.lanl.gov/abs/hep-th/#1}{arXiv:\allowbreak
hep-th\slash{\xold#1}}}

\def\arxiv#1#2{\href{http://arxiv.org/abs/#1.#2}{arXiv:\allowbreak
{\xold#1}.{\xold#2}}} 
\def\jhep#1#2#3#4{\href{http://jhep.sissa.it/stdsearch?paper=#2\%28#3\%29#4}{J. High Energy Phys. {\xbold #1#2} ({\xold#3}) {\xold#4}}}

\def\CMP#1#2#3{Commun. Math. Phys. {\xbold#1} ({\xold#2}) {\xold#3}}
\def\CQG#1#2#3{Class. Quantum Grav. {\xbold#1} ({\xold#2}) {\xold#3}}

\def\IJMPA#1#2#3{Int. J. Mod. Phys. {\xbf A}{\xbold#1} ({\xold#2}) {\xold#3}}

\def\JHEP{\jhep}

\def\JPA#1#2#3{J. Phys. {\xbf A}{\xbold#1} ({\xold#2}) {\xold#3}}
\def\LMP#1#2#3{Lett. Math. Phys. {\xbold#1} ({\xold#2}) {\xold#3}}
\def\MPLA#1#2#3{Mod. Phys. Lett. {\xbf A}{\xbold#1} ({\xold#2}) {\xold#3}}

\def\NPB#1#2#3{Nucl. Phys. {\xbf B}{\xbold#1} ({\xold#2}) {\xold#3}}

\def\PLB#1#2#3{Phys. Lett. {\xbf B}{\xbold#1} ({\xold#2}) {\xold#3}}
\def\PR#1#2#3{Phys. Rept. {\xbold#1} ({\xold#2}) {\xold#3}}
\def\PRD#1#2#3{Phys. Rev. {\xbf D}{\xbold#1} ({\xold#2}) {\xold#3}}
\def\PRL#1#2#3{Phys. Rev. Lett. {\xbold#1} ({\xold#2}) {\xold#3}}

\newcount\sectioncount
\sectioncount=0
\def\section#1#2{\global\eqcount=0
	\global\subsectioncount=0
        \global\advance\sectioncount by 1
	\ifnum\sectioncount>1
	        \vskip2\baselineskip
	\fi
\line{\twelvecp\the\sectioncount. #2\hfill}
       \vskip.5\baselineskip\noindent
        \xdef#1{{\old\the\sectioncount}}}
\newcount\subsectioncount
\def\subsection#1#2{\global\advance\subsectioncount by 1
\vskip.75\baselineskip\noindent\line{\tencp\the\sectioncount.\the\subsectioncount. #2\hfill}\nobreak\vskip.4\baselineskip\nobreak\noindent\xdef#1{{\old\the\sectioncount}.{\old\the\subsectioncount}}}
\def\immediatesubsection#1#2{\global\advance\subsectioncount by 1
\vskip-\baselineskip\noindent
\line{\tencp\the\sectioncount.\the\subsectioncount. #2\hfill}
	\vskip.5\baselineskip\noindent
	\xdef#1{{\old\the\sectioncount}.{\old\the\subsectioncount}}}
\newcount\subsubsectioncount
\def\subsubsection#1#2{\global\advance\subsubsectioncount by 1
\vskip.75\baselineskip\noindent\line{\tencp\the\sectioncount.\the\subsectioncount.\the\subsubsectioncount. #2\hfill}\nobreak\vskip.4\baselineskip\nobreak\noindent\xdef#1{{\old\the\sectioncount}.{\old\the\subsectioncount}.{\old\the\subsubsectioncount}}}
\newcount\appendixcount
\appendixcount=0
\def\appendix#1{\global\eqcount=0
        \global\advance\appendixcount by 1
        \vskip2\baselineskip\noindent
        \ifnum\the\appendixcount=1
        \hbox{\twelvecp Appendix A: #1\hfill}\vskip\baselineskip\noindent\fi
    \ifnum\the\appendixcount=2
        \hbox{\twelvecp Appendix B: #1\hfill}\vskip\baselineskip\noindent\fi
    \ifnum\the\appendixcount=3
        \hbox{\twelvecp Appendix C: #1\hfill}\vskip\baselineskip\noindent\fi}
\def\acknowledgements{\vskip2\baselineskip\noindent
        \underbar{\it Acknowledgements:}\ }
\newcount\eqcount
\eqcount=0
\def\Eqn#1{\global\advance\eqcount by 1
\ifnum\the\sectioncount=0
	\xdef#1{{\noexpand\oldsize\the\eqcount}}
	\eqno({\oldstyle\the\eqcount})
\else
        \ifnum\the\appendixcount=0
\xdef#1{{\noexpand\oldsize\the\sectioncount}.{\noexpand\oldsize\the\eqcount}}
                \eqno({\oldstyle\the\sectioncount}.{\oldstyle\the\eqcount})\fi
        \ifnum\the\appendixcount=1
	        \xdef#1{{\noexpand\oldstyle A}.{\noexpand\oldstyle\the\eqcount}}
                \eqno({\oldstyle A}.{\oldstyle\the\eqcount})\fi
        \ifnum\the\appendixcount=2
	        \xdef#1{{\noexpand\oldstyle B}.{\noexpand\oldstyle\the\eqcount}}
                \eqno({\oldstyle B}.{\oldstyle\the\eqcount})\fi
        \ifnum\the\appendixcount=3
	        \xdef#1{{\noexpand\oldstyle C}.{\noexpand\oldstyle\the\eqcount}}
                \eqno({\oldstyle C}.{\oldstyle\the\eqcount})\fi
\fi}
\def\eqn{\global\advance\eqcount by 1
\ifnum\the\sectioncount=0
	\eqno({\oldstyle\the\eqcount})
\else
        \ifnum\the\appendixcount=0
                \eqno({\oldstyle\the\sectioncount}.{\oldstyle\the\eqcount})\fi
        \ifnum\the\appendixcount=1
                \eqno({\oldstyle A}.{\oldstyle\the\eqcount})\fi
        \ifnum\the\appendixcount=2
                \eqno({\oldstyle B}.{\oldstyle\the\eqcount})\fi
        \ifnum\the\appendixcount=3
                \eqno({\oldstyle C}.{\oldstyle\the\eqcount})\fi
\fi}
\def\multi{\global\advance\eqcount by 1}
\def\multieqn#1{({\oldstyle\the\sectioncount}.{\oldstyle\the\eqcount}\hbox{#1})}
\def\multiEqn#1#2{\xdef#1{{\oldstyle\the\sectioncount}.{\old\the\eqcount}#2}
        ({\oldstyle\the\sectioncount}.{\oldstyle\the\eqcount}\hbox{#2})}
\def\multiEqnAll#1{\xdef#1{{\oldstyle\the\sectioncount}.{\old\the\eqcount}}}
\newcount\tablecount
\tablecount=0
\def\Table#1#2{\global\advance\tablecount by 1
       \xdef#1{\the\tablecount}
       \vskip2\parskip
       \centerline{\it Table \the\tablecount: #2}
       \vskip\parskip}
\newtoks\url
\def\Href#1#2{\catcode`\#=12\url={#1}\catcode`\#=\active#2}
\def\href#1#2{{#2}}

\parskip=3.5pt plus .3pt minus .3pt
\baselineskip=14pt plus .1pt minus .05pt
\lineskip=.5pt plus .05pt minus .05pt
\lineskiplimit=.5pt
\abovedisplayskip=18pt plus 4pt minus 2pt
\belowdisplayskip=\abovedisplayskip
\hsize=14cm
\vsize=19cm
\hoffset=1.5cm
\voffset=1.8cm
\frenchspacing
\footline={}
\raggedbottom

\newskip\origparindent
\origparindent=\parindent

\def\sss{\scriptscriptstyle}
\def\*{\partial}
\def\punkt{\,\,.}
\def\komma{\,\,,}

\def\={\!=\!}
\def\small#1{{\hbox{$#1$}}}

\def\fraction#1{\small{1\over#1}}
\def\fr{\fraction}
\def\Fraction#1#2{\small{#1\over#2}}
\def\Fr{\Fraction}

\def\eg{{\it e.g.}}

\def\ie{{\it i.e.}}

\def\nlni{\hfill\break}

\def\a{\alpha}
\def\b{\beta}
\def\d{\delta}
\def\e{\varepsilon}
\def\g{\gamma}
\def\l{\lambda}

\def\L{\Lambda}

\def\RR{{\Bbb R}}


\def\L{\Lambda}

\def\textfrac#1#2{\raise .45ex\hbox{\the\scriptfont0 #1}\nobreak\hskip-1pt/\hskip-1pt\hbox{\the\scriptfont0 #2}}

\def\LL{{\cal L}}
\def\leftbr{[\![}
\def\rightbr{]\!]}
\def\leftpar{(\!(}
\def\rightpar{)\!)}


\def\frac{\Fr}

\def\mathbb{\Bbb}
\def\cd{\!\cdot\!}


\ref\CederwallUfoldbranes{M. Cederwall, {\xit ``M-branes on U-folds''},
in proceedings of 7th International Workshop ``Supersymmetries and
Quantum Symmetries'' Dubna, 2007 [\arxiv{0712}{4287}].}

\ref\BermanPerryGen{D.S. Berman and M.J. Perry, {\xit ``Generalised
geometry and M-theory''}, \jhep{11}{06}{2011}{074} [\arxiv{1008}{1763}].}    

\ref\BermanMusaevThompson{D.S. Berman, E.T. Musaev and D.C. Thompson,
{\xit ``Duality invariant M-theory: gaugings via Scherk--Schwarz
reduction}, \arxiv{1208}{0020}.}

\ref\UdualityMembranes{V. Bengtsson, M. Cederwall, H. Larsson and
B.E.W. Nilsson, {\xit ``U-duality covariant
membranes''}, \jhep{05}{02}{2005}{020} [\hepth{0406223}].}

\ref\ObersPiolineU{N.A. Obers and B. Pioline, {\xit ``U-duality and M-theory''},
\PR{318}{1999}{113}, 
\nlni [\hepth{9809039}].}

\ref\BermanGodazgarPerry{D.S. Berman, H. Godazgar and M.J. Perry,
{\xit ``SO(5,5) duality in M-theory and generalized geometry''},
\arxiv{1103}{5733}.} 

\ref\BermanMusaevPerry{D.S. Berman, E.T. Musaev and M.J. Perry,
{\xit ``Boundary terms in generalized geometry and doubled field theory''},
\arxiv{1110}{3097}.} 

\ref\BermanGodazgarGodazgarPerry{D.S. Berman, H. Godazgar, M. Godazgar  
and M.J. Perry,
{\xit ``The local symmetries of M-theory and their formulation in
generalised geometry''}, 
\arxiv{1110}{3930}.} 

\ref\BermanGodazgarPerryWest{D.S. Berman, H. Godazgar, M.J. Perry and
P. West,
{\xit ``Duality invariant actions and generalised geometry''}, 
\arxiv{1111}{0459}.} 

\ref\CoimbraStricklandWaldram{A. Coimbra, C. Strickland-Constable and
D. Waldram, {\xit ``$E_{d(d)}\times\hbox{\eightbbb R}^+$ generalised geometry,
connections and M theory'' }, \arxiv{1112}{3989}.} 

\ref\CremmerPopeI{E. Cremmer, B. Julia, H. L\"u and C.N. Pope,
{\xit ``Dualisation of dualities. I.''}, \NPB{523}{1998}{73} [\hepth{9710119}].}

\ref\HullT{C.M. Hull, {\xit ``A geometry for non-geometric string
backgrounds''}, \jhep{05}{10}{2005}{065} [\hepth{0406102}].}

\ref\HullM{C.M. Hull, {\xit ``Generalised geometry for M-theory''},
\jhep{07}{07}{2007}{079} [\hepth{0701203}].}

\ref\HullDoubled{C.M. Hull, {\xit ``Doubled geometry and
T-folds''}, \jhep{07}{07}{2007}{080}
[\hepth{0605149}].}

\ref\HullTownsend{C.M. Hull and P.K. Townsend, {\xit ``Unity of
superstring dualities''}, \NPB{438}{1995}{109} [\hepth{9410167}].}

\ref\PalmkvistHierarchy{J. Palmkvist, {\xit ``Tensor hierarchies,
Borcherds algebras and $E_{11}$''}, \arxiv{1110}{4892}.}

\ref\deWitNicolaiSamtleben{B. de Wit, H. Nicolai and H. Samtleben,
{\xit ``Gauged supergravities, tensor hierarchies, and M-theory''},
\jhep{02}{08}{2008}{044} [\arxiv{0801}{1294}].}

\ref\deWitSamtleben{B. de Wit and H. Samtleben,
{\xit ``The end of the $p$-form hierarchy''},
\jhep{08}{08}{2008}{015} [\arxiv{0805}{4767}].}

\ref\CederwallJordanMech{M.~Cederwall, {\xit ``Jordan algebra
dynamics''}, \PLB{210}{1988}{169}.} 

\ref\BerkovitsNekrasovCharacter{N. Berkovits and N. Nekrasov, {\xit
    ``The character of pure spinors''}, \LMP{74}{2005}{75}
  [\hepth{0503075}].}

\ref\HitchinLectures{N. Hitchin, {``\xit Lectures on generalized
geometry''}, \arxiv{1010}{2526}.}

\ref\KoepsellNicolaiSamtleben{K. Koepsell, H. Nicolai and
H. Samtleben, {\xit ``On the Yangian $[Y(e_8)]$ quantum symmetry of
maximal supergravity in two dimensions''}, \jhep{99}{04}{1999}{023}
[\hepth{9903111}].}

\ref\HohmHullZwiebachI{O. Hohm, C.M. Hull and B. Zwiebach, {\xit ``Background
independent action for double field
theory''}, \jhep{10}{07}{2010}{016} [\arxiv{1003}{5027}].}

\ref\HohmHullZwiebachII{O. Hohm, C.M. Hull and B. Zwiebach, {\xit
``Generalized metric formulation of double field theory''},
\jhep{10}{08}{2010}{008} [\arxiv{1006}{4823}].} 

\ref\HohmZwiebach{O. Hohm and B. Zwiebach, {\xit ``On the Riemann
tensor in double field theory''}, \jhep{12}{05}{2012}{126}
[\arxiv{1112}{5296}].} 

\ref\WestEEleven{P. West, {\xit ``$E_{11}$ and M theory''},
\CQG{18}{2001}{4443} [\hepth{0104081}].}

\ref\AndriotLarforsLustPatalong{D. Andriot, M. Larfors, D. L\"ust and
P. Patalong, {\xit ``A ten-dimensional action for non-geometric
fluxes''}, \jhep{11}{09}{2011}{134} [\arxiv{1106}{4015}].}

\ref\AndriotHohmLarforsLustPatalongI{D. Andriot, O. Hohm, M. Larfors,
D. L\"ust and 
P. Patalong, {\xit ``A geometric action for non-geometric
fluxes''}, \PRL{108}{2012}{261602} [\arxiv{1202}{3060}].}

\ref\AndriotHohmLarforsLustPatalongII{D. Andriot, O. Hohm, M. Larfors,
D. L\"ust and 
P. Patalong, {\xit ``Non-geometric fluxes in supergravity and double
field theory''}, \arxiv{1204}{1979}].}

\ref\DamourHenneauxNicolai{T. Damour, M. Henneaux and H. Nicolai,
{\xit ``Cosmological billiards''}, \CQG{20}{2003}{R145} [\hepth{0212256}].}

\ref\DamourNicolai{T. Damour and H. Nicolai, 
{\xit ``Symmetries, singularities and the de-emergence of space''},
\arxiv{0705}{2643}.}

\ref\EHTP{F. Englert, L. Houart, A. Taormina and P. West,
{\xit ``The symmetry of M theories''},
\jhep{03}{09}{2003}{020}2003 [\hepth{0304206}].}

\ref\PachecoWaldram{P.P. Pacheco and D. Waldram, {\xit ``M-theory,
exceptional generalised geometry and superpotentials''},
\jhep{08}{09}{2008}{123} [\arxiv{0804}{1362}].}

\ref\DamourHenneauxNicolaiII{T. Damour, M. Henneaux and H. Nicolai,
{\xit ``$E_{10}$ and a 'small tension expansion' of M theory''},
\PRL{89}{2002}{221601} [\hepth{0207267}].}

\ref\KleinschmidtNicolai{A. Kleinschmidt and H. Nicolai, {\xit
``$E_{10}$ and $SO(9,9)$ invariant supergravity''},
\jhep{04}{07}{2004}{041} [\hepth{0407101}].}

\ref\WestII{P.C. West, {\xit ``$E_{11}$, $SL(32)$ and central charges''},
\PLB{575}{2003}{333} [\hepth{0307098}].}

\ref\KleinschmidtWest{A. Kleinschmidt and P.C. West, {\xit
``Representations of $G^{+++}$ and the r\^ole of space-time''},
\jhep{04}{02}{2004}{033} [\hepth{0312247}].}

\ref\WestIII{P.C. West, {\xit ``$E_{11}$ origin of brane charges and
U-duality multiplets''}, \jhep{04}{08}{2004}{052} [\hepth{0406150}].}

\ref\PiolineWaldron{B. Pioline and A. Waldron, {\xit ``The automorphic
membrane''}, \jhep{04}{06}{2004}{009} [\hepth{0404018}].}

\ref\WestBPS{P.C. West, {\xit ``Generalised BPS conditions''},
\arxiv{1208}{3397}.} 

\ref\JeonLeeParkI{I. Jeon, K. Lee and J.-H. Park, {\xit ``Differential
geometry with a projection: Application to double field theory''},
\jhep{11}{04}{2011}{014} [\arxiv{1011}{1324}].}

\ref\JeonLeeParkII{I. Jeon, K. Lee and J.-H. Park, {\xit ``Stringy
differential geometry, beyond Riemann''}, 
\PRD{84}{2011}{044022} [\arxiv{1105}{6294}].}

\ref\JeonLeeParkIII{I. Jeon, K. Lee and J.-H. Park, {\xit
``Supersymmetric double field theory: stringy reformulation of supergravity''},
\PRD{85}{2012}{081501} [\arxiv{1112}{0069}].}

\ref\Hillmann{C. Hillmann, {\xit ``Generalized $E_{7(7)}$ coset
dynamics and $D=11$ supergravity''}, \jhep{09}{03}{2009}{135}
[\arxiv{0901}{1581}].}

\ref\CoimbraStricklandWaldramSG{A. Coimbra, C. Strickland-Constable and
D. Waldram, {\xit ``Generalised geometry and type II supergravity''}, 
\arxiv{1202}{3170}.} 

\ref\JeonLeeParkFermions{I. Jeon, K. Lee and J.-H. Park, {\xit
``Incorporation of fermions into double field theory''}, 
\jhep{11}{11}{2011}{025} [\arxiv{1109}{2035}].}

\ref\KleinschmidtCounting{A. Kleinschmidt, {\xit ``Counting of
supersymmetric branes''}, \arxiv{1109}{2025}.}

\ref\HohmKwak{O. Hohm and S.K. Kwak, {\xit ``$N=1$ supersymmetric
double field theory''}, \jhep{12}{03}{2012}{080} [\arxiv{1111}{7293}].}

\ref\CremmerLuPopeStelle{E. Cremmer, H. L\"u, C.N. Pope and
K.S. Stelle, {\xit ``Spectrum-generating symmetries for BPS solitons''},
\NPB{520}{1998}{132} [\hepth{9707207}].}

\ref\deWitNicolai{B. de Wit and H. Nicolai, {\xit ``$D = 11$
supergravity with local $SU(8)$ invariance''}, \NPB{274}{1986}{363}.}

\ref\deWitMBPS{B.de Wit,
 {\xit ``M theory duality and BPS extended supergravity''},
 \IJMPA{16}{2001}{1002} [\hepth{0010292}].} 

\ref\WestHidden{P. West, {\xit ``Hidden superconformal symmetry in M
theory''}, \JHEP{00}{08}{2000}{007} [\hepth{0005270}].}

\ref\KoepsellNicolaiSamtlebenExc{K. Koepsell, H. Nicolai and
H. Samtleben, {\xit ``An exceptional geometry for $D = 11$
supergravity?''}, \CQG{17}{2000}{368} [\hepth{0006034}].} 

\ref\deWitNicolaiAllThat{B. de Wit and H. Nicolai,
 {\xit ``Hidden symmetries, central charges and all that''}, 
\CQG{18}{2001}{3095} [\hepth{0011239}].}

\ref\JuliaNuffield{B. Julia, {\xit ``Group disintegrations''}, 
Invited paper presented at Nuffield
Gravity Workshop, Cambridge, England, Jun {\xold22}--Jul {\xold12}, LPT-ENS
{\xold80}-{\xold16} ({\xold1980}).}

\ref\JuliaLectures{B. Julia, {\xit ``Kac-Moody symmetry of gravitation and supergravity theories''}, in Lectures in Applied Mathematics,
AMS-SIAM {\xbold21} ({\xold1985}) {\xold355} [LPT-ENS {\xold82}-{\xold22}].}

\ref\NicolaiIntegrability{H. Nicolai,
 {\xit ``The integrability of $N=16$ supergravity''}, \PLB{194}{1987}{402}.}

\ref\GunaydinConformal{M. G\"unaydin, {\xit ``Generalized conformal and
superconformal group actions and Jordan algebras''},
\MPLA8{1993}{1407} [\hepth{9301050}].} 

\ref\GunaydinKoepsellNicolai{M G\"unaydin, K. Koepsell and H. Nicolai,
{\xit ``Conformal and quasiconformal realizations of exceptional Lie
groups''}, \CMP{221}{2001}{57} [\hepth{00068083}].}

\ref\DamourKleinschmidtNicolai{T. Damour, A. Kleinschmidt and
H. Nicolai, {\xit ``$K(E_{10})$, supergravity and fermions''},
\jhep{06}{08}{2006}{046} [\hepth{0606105}].}

\ref\HenryJuliaPaulot{P. Henry-Labord\`ere, B.L. Julia and L. Paulot,
{\xit ``Borcherds symmetries in M-theory''}, \jhep{02}{04}{2002}{049}
[\hepth{0203070}].} 

\ref\HennauxJuliaLevie{M. Henneaux, B.L. Julia and J. Levie, {\xit
``$E_{11}$, Borcherds algebras and maximal supergravity''},
\jhep{12}{04}{2012}{078} [\arxiv{1007}{5241}].}

\ref\HohmKwakFrame{O. Hohm and S.K. Kwak, {\xit ``Frame-like geometry
of double field theory''}, \JPA{44}{2011}{085404} [\arxiv{1011}{4101}].}

\ref\HohmKwakZwiebachI{O. Hohm, S.K. Kwak and B. Zwiebach, {\xit
``Unification of type II strings and T-duality''},
\PRL{107}{2011}{171603} [\arxiv{1106}{5452}].}

\ref\HohmKwakZwiebachII{O. Hohm, S.K. Kwak and B. Zwiebach, {\xit
``Double field theory of type II strings''}, \jhep{11}{09}{2011}{013}
[\arxiv{1107}{0008}].} 

\ref\GreenMillerVanhove{M.B. Green, S.D. Miller and P. Vanhove, {\xit
``Small representations, string instantons, and Fourier modes of
Eisenstein series''}, \arxiv{1111}{2983}.}


\line{
\epsfysize=9mm
\epsffile{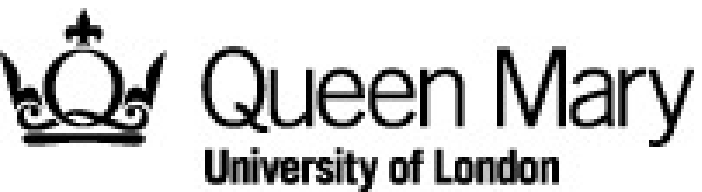}
\hskip3mm\epsfysize=9mm
\epsffile{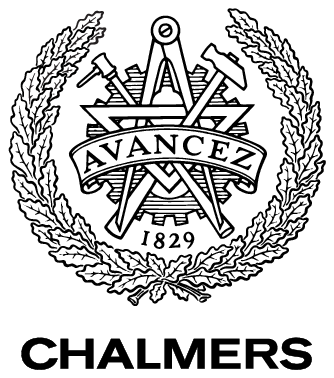}
\hskip3mm\epsfysize=9mm
\epsffile{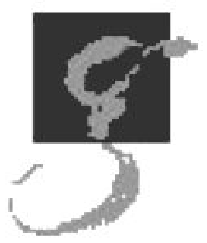}
\hskip3mm\epsfysize=9mm
\epsffile{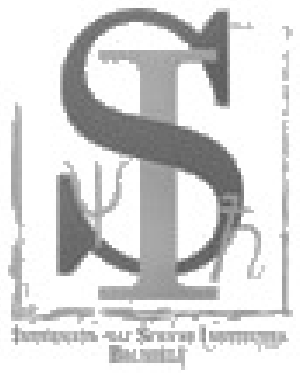}
\hskip3mm\epsfysize=9mm
\epsffile{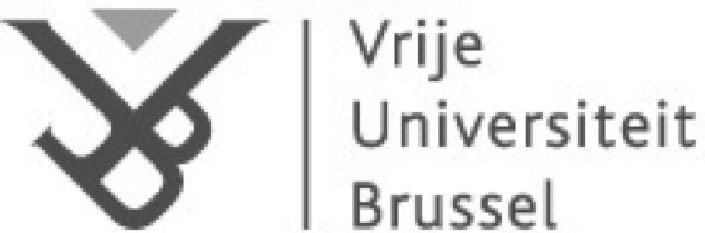}
\hfill}
\vskip-16mm
\vtop{\baselineskip=.75\baselineskip
\line{\hfill\xrm QMUL-PH-12-14}
\line{\hfill\xrm Gothenburg preprint}
\line{\hfill\xrm AEI-2012-085}
\line{\hfill\xrm August, {\old2012}}
}

\vskip\parskip
\line{\hrulefill}

\headtext={Berman, Cederwall, Kleinschmidt, Thompson: 
``The gauge structure...''}

\vfill
\vskip.5cm

\centerline{\sixteenhelvbold
The gauge structure of generalised diffeomorphisms}

\vfill

\centerline{\twelvehelvbold{David S. Berman${}^1$, Martin Cederwall${}^2$,}}

\vskip\parskip

\centerline{\twelvehelvbold{Axel Kleinschmidt${}^{3,4}$ 
and Daniel C. Thompson${}^{4,5}$}}

\vskip.8cm

\centerline{${}^{\sss1}$\xit Centre for Research in String Theory,
School of Physics, Queen Mary College,}
\vskip-1mm
\centerline{\xit University of London, Mile End Road, 
London, E1 4NS, England}

\vskip\parskip

\centerline{${}^{\sss2}$\xit Dept. of Fundamental Physics,
Chalmers University of Technology,
SE 412 96 Gothenburg, Sweden}

\vskip\parskip

\centerline{${}^{\sss3}$\xit Max-Planck-Institut f\"ur Gravitationsphysik 
(Albert Einstein Institut),}
\vskip-1mm
\centerline{\xit Am M\"uhlenberg 1, 14476 Golm, Germany}

\vskip\parskip

\centerline{${}^{\sss4}$\xit International Solvay Institutes, Campus
Plaine C.P. 231,} 
\vskip-1mm
\centerline{\xit Boulevard du Triomphe, 1050 Bruxelles, Belgium}

\vskip\parskip

\centerline{${}^{\sss5}$\xit Theoretische Natuurkunde, Vrije
Universiteit Brussel,}  
\vskip-1mm
\centerline{\xit International Solvay Institutes,
Pleinlaan 2, B-1050, Brussels, Belgium}

\vfill

{\narrower\noindent \underbar{Abstract:} We investigate the
generalised diffeomorphisms in M-theory, which are gauge
transformations unifying diffeomorphisms and tensor gauge
transformations.
After giving an $E_{n(n)}$-covariant description of the gauge
transformations and their commutators, we show that the gauge algebra
is infinitely reducible, \ie, the tower of ghosts for ghosts is
infinite. The Jacobiator of generalised diffeomorphisms gives such a
reducibility transformation. 
We give a concrete description of the ghost structure, and
demonstrate that the infinite sums give the correct (regularised)
number of degrees of freedom. The ghost towers belong to the
sequences of representations previously observed appearing in tensor
hierarchies and Borcherds algebras.
All calculations rely on the section condition, which we reformulate
as a linear condition on the cotangent directions.
The analysis holds for $n<8$. At $n=8$, where the dual gravity field becomes
relevant, the natural guess for the gauge parameter and its
reducibility still yields the correct counting of gauge parameters.
\smallskip}
\vfill

\font\xxtt=cmtt6

\vtop{\baselineskip=.6\baselineskip\xxtt
\line{\hrulefill}
\catcode`\@=11
\line{email: d.s.berman@qmul.ac.uk, martin.cederwall@chalmers.se, 
Axel.Kleinschmidt@aei.mpg.de, dthompson@tena4.vub.ac.be\hfill}
\catcode`\@=\active
}

\eject

\section\Introduction{Introduction}It has been known for a long time
that compactification on an $n$-dimensional torus of $D=11$
supergravity, and of M-theory, 
enjoys a U-duality symmetry, namely a discrete version of $E_{n(n)}$
(see for example refs. [\HullTownsend,\CremmerPopeI,\ObersPiolineU]). Such a
symmetry mixes momentum states with winding states of branes. 
The series can even be continued to the infinite-dimensional
algebras 
$E_9$ [\JuliaNuffield,\NicolaiIntegrability], 
$E_{10}$ [\JuliaLectures\skipref\DamourHenneauxNicolai\skipref\DamourNicolai\skipref\DamourHenneauxNicolaiII-\KleinschmidtNicolai] 
and $E_{11}$ 
[\WestEEleven\skipref\WestII\skipref\KleinschmidtWest\skipref\WestIII-\EHTP], 
although the interpretation is
somewhat different in the last two cases.
It has
later become clear that it should be possible to give the theory a
formulation which is manifest under the (continuous) exceptional
group 
[\deWitNicolai\skipref\deWitMBPS\skipref\WestHidden\skipref\KoepsellNicolaiSamtlebenExc\skipref\deWitNicolaiAllThat-\Hillmann],
which plays roughly speaking the same r\^ole as 
$GL(n)$ does in gravity. 

Such ideas, where space-time is enlarged to accommodate
the extra ``momenta'', found its
geometric formulation in the work of Hull, first
for T-duality [\HullT] and later for U-duality [\HullM,\PachecoWaldram].
The doubled field theory relevant for T-duality is closely connected
to the generalised geometry of Hitchin [\HitchinLectures], and has
been thoroughly investigated
[\HullDoubled\skipref\HohmHullZwiebachI\skipref\HohmHullZwiebachII\skipref\HohmZwiebach\skipref\AndriotLarforsLustPatalong\skipref\AndriotHohmLarforsLustPatalongI\skipref\AndriotHohmLarforsLustPatalongII\skipref\JeonLeeParkI\skipref\JeonLeeParkII\skipref\JeonLeeParkIII-\HohmKwakFrame],
although some geometric understanding still seems to be missing.

Concerning U-duality, much of the structure is similar, but there are
some fundamental differences in the structure of the generalised
diffeomorphisms. Part of the purpose of the present paper is to
clarify these.
From investigations of the dynamics of supersymmetric membranes, it
has been clear from different arguments that U-duality can be made
manifest
[\CederwallUfoldbranes,\BermanPerryGen,\UdualityMembranes]. Properties
of generalised diffeomorphisms, and of some objects transforming
linearly under them (that can serve as equations of motion) have been
investigated in
refs. [\BermanMusaevThompson\skipref\BermanGodazgarPerry\skipref\BermanMusaevPerry\skipref\BermanGodazgarGodazgarPerry\skipref\BermanGodazgarPerryWest-\CoimbraStricklandWaldram]. 
A generic geometric picture and a tensor formalism are lacking, as is
a full geometric treatment of fermions (though progress has been made
in the case of double field theory/ten-dimensional supergravity
[\HohmKwakZwiebachI\skipref\HohmKwakZwiebachII\skipref\CoimbraStricklandWaldramSG\skipref\JeonLeeParkFermions-\HohmKwak]).

An interesting structure arising in the context of U-duality 
is the concept of tensor hierarchies, connected
to the possible gaugings of supergravity
[\deWitNicolaiSamtleben,\deWitSamtleben,\KleinschmidtCounting]. They can
also be understood in terms of Borcherds algebras
[\HenryJuliaPaulot,\HennauxJuliaLevie,\PalmkvistHierarchy], 
and seem to have an origin in the decomposition
of the adjoint representation of $E_{11}$ [\WestIII,\WestBPS].
In the present paper we will encounter the same structures in a
seemingly different context, namely in the ghost structure of the
algebra of generalised diffeomorphisms. Some of the representations
are given in Table 1. 
The representations $R_k$ given there are possible
representations of 
$k$-form fields in the uncompactified dimensions. 

\vskip4\parskip
\ruledtable
 $n$ |  $R_1$ & $R_2$ & $R_3$ & $R_4$  \cr
3 |    $({\bf3},{\bf2})$   &  $(\overline{\bf3},{\bf1})$  &
$({\bf1},{\bf2})$  &   $({\bf3},{\bf1})$  \cr 
4 | ${\bf10}$& $\overline{\bf5}$ & ${\bf5}$&
 $\overline{\bf10}$ \cr   
5 |  $\bf16$ & $\bf10$ & $\overline{\bf16}$ & $\bf45$ \cr 
6 |  $\bf27$ & $\overline{\bf27}$ & $\bf78$ & $\overline{\bf351'}$ \cr
7 | $\bf{56}$ & $\bf{133}$  & $\bf{912}$  &
${\bf8645}\oplus{\bf133}$ 
\cr
8 | $\bf{248}$ &  ${\bf3875}\oplus{\bf1}$ 
& ${\bf147250}\oplus{\bf3875}\oplus{\bf248}$ 
& ${\bf6696000}\oplus{\bf779247}\oplus{\bf147250}
$\nr
|&&&${}\oplus2\cd{\bf30380}
  \oplus{\bf3875}\oplus2\cd{\bf248}$
 \endruledtable
\Table\ReducibilityTable{Some relevant representations}

The section condition, restricting the dependence of fields on the
coordinates of the extended manifold, is central in the analysis. It
is essential for the equivalence of a model formulated within the
framework of generalised geometry with an ordinary supergravity. If it
could be meaningfully continued to higher $n$, it would be a key
ingredient in demonstrating the validity of the $E_{11}$ conjecture. A
first step in this direction may be found in ref. [\WestBPS].

The present paper is structured as follows: Section 2 contains basic
facts about generalised (exceptional) diffeomorphisms and their
algebras. This analysis is not new [\CoimbraStricklandWaldram], 
but its covariant formulation has
to our knowledge not been given previously. In Section 3, we
investigate the reducibility of generalised diffeomorphisms, which,
when formulated covariantly, turns out to be infinite. The version of
the representations presented in Table \ReducibilityTable\ is part of
the prediction of Section 3. We show that
the counting of gauge parameters arising from the infinite sums is
correct, even for $n=8$. In Section 4, finally, we come back to the
section condition and show how it can be cast in a linear form,
mimicking the construction of isotropic subspaces from pure spinors in
doubled geometry [\HitchinLectures].
Throughout the paper we will only be concerned with what is normally
viewed as the compactified directions, and completely ignore the rest.

A note on terminology: The term ``generalised geometry'', when used in
this paper, will typically refer to the exceptional $E_{n(n)}$
geometry. When we want to refer to the $O(d,d)$ situation, we use the
term ``doubled geometry''. We have no need to make a terminological
distinction between the doubled formalism of Hull [\HullT] and the
mathematical setting of Hitchin [\HitchinLectures].

\vfill\eject
\section\GeneralisedDiffeos{Generalised diffeomorphisms}Since
gravitational and tensorial degrees of freedom mix under 
U-duality, so do their respective gauge transformations, and the concept
of diffeomorphisms has to be generalised. In ordinary geometry, the
infinitesimal transformation of any
tensor is given by the Lie derivative in the direction of the
diffeomorphism parameter $u^m$, which acting on a vector $v^m$ reads
$$
\d_u v^m=L_u v^m=[u,v]^m=u^n\*_n v^m-\*_n u^m v^n\punkt\eqn
$$
The interpretation of this transformation that best lends itself to
generalisations is to view the first term as a transport term, and the
second one as a ${\frak gl}(n)$ transformation with the matrix $\*_nu^m$
valued in the fundamental representation of the Lie algebra ${\frak gl}(n)$. 
The transformation of any tensor is given by replacing the Lie algebra
action by the appropriate representation.
Of course, this transformation is
already antisymmetric in $u$ and $v$, and the commutator of two
diffeomorphisms is given by the algebra of vector fields:
$$
[L_u,L_v]=L_{[u,v]}=L_{L_u v}\punkt\eqn
$$

In the context of U-duality, the r\^ole played by ${\frak gl}(n)$ is assumed
by the Lie algebra ${\frak e}_{n(n)}$, together with a real
scaling, corresponding to the (on-shell) 
trombone symmetry [\CremmerLuPopeStelle]. 
The tensors should now be tensors under $E_{n(n)}\times\RR$,
and a generalised diffeomorphism
should be of the form
$$
\d_UV^M=\LL_UV^M=U^N\*_NV^M
-\a P_{(adj)}{}^M{}_{N,}{}^P{}_Q\*_PU^QV^N+\b\*_NU^NV^M\Eqn\AlgebraTransf
$$
for some constants $\a$ and $\beta$.
An upper index $M,N,\ldots$ denotes an object in the representation
$R_1$ (see Table \ReducibilityTable), 
the ``coordinate representation'' of $E_{n(n)}$, 
and $P_{(adj)}$ is the projection on the adjoint
representation that is contained in the tensor product 
$R_1\otimes \bar{R}_1$ of the coordinate representation $R_1$ and its
conjugate.
This lends itself to a natural generalisation to other
representations, and respects composition of tensors since both terms
obey the Leibniz rule.
These transformations has been been defined, and their algebra examined, in
ref. [\CoimbraStricklandWaldram], in a formalism where $GL(n)$ is
manifest. 
We will give a covariant analysis.
The operation $\LL_U$ is referred to as a generalised Lie derivative,
or an ``exceptional Dorfman bracket''.

The Ansatz we will use is not of the form (\AlgebraTransf), but of the
general form
$$
\LL_UV^M=L_UV^M+Y^{MN}{}_{PQ}\*_NU^PV^Q\Eqn\LieDerAnsatz
$$
for some $E_{n(n)}$-invariant tensor $Y$. If the ``tensor-friendly''
version (\AlgebraTransf) is to hold, there must be an identity
$$
Y^{MN}{}_{PQ}=\d^M_P\d^N_Q
-\a P_{(adj)}{}^M{}_{Q,}{}^N{}_P+\b\d^M_Q\d^N_P\punkt\eqn
$$

\subsection\TheSectionCondition{The section condition}When one starts 
commuting two generalised diffeomorphisms 
given by eq. (\LieDerAnsatz), one immediately
encounters the condition 
$$
Y^{MN}{}_{PQ}\*_M\ldots\*_N\ldots=0\komma\eqn
$$
where the ellipses indicate that the derivatives act on different
objects. This will be solved by the {\it section condition}. Often, one
makes the difference between a {\it strong} and a {\it weak} section
condition. The weak version of the section condition reads
$P_{(R_2)}^{MN}{}_{PQ}\*_M\*_N\Phi=0$ for any field or variable $\Phi$,
while the strong one states that
$P_{(R_2)}^{MN}{}_{PQ}\*_M\Phi\*_N\Phi'=0$ for any pair of fields or
variables, which we write in shorthand as
$$
(\*\otimes\*)|_{\bar R_2}=0\punkt\Eqn\ExplSectCond
$$
(The representation $R_2$ is listed for the various values of $n$ in
Table \ReducibilityTable.)
 This latter version is the one that will be needed
throughout our analysis, and when we refer to the section condition in
the following, we will mean its strong version, unless explicitly
stated otherwise.

Section 4 will be devoted to a detailed analysis of the section
condition, its reformulation and solution. For now, we just make
a brief comment. The interpretation of eq. (\ExplSectCond) is not as a
non-linear condition on the momenta (which would be strange,
considering they are cotangent vectors). Instead, eq. (\ExplSectCond)
should be read: Find a largest possible linear subspace of cotangent
space such that any pair of vectors $A$ and $B$ belonging to the
subspace fulfill $P_{(R_2)}^{MN}{}_{PQ}A_MB_N=0$. This insinuates
that there should be a more direct way of writing the section
condition as a linear condition on momenta. We will do this in 
Section 4. Generally, any such solution to the section
condition will pick out an $n$-dimensional subspace conserved by $GL(n)$.

In doubled geometry, where the manifest duality group is $O(d,d)$,
the generalised Lie derivative (Dorfman bracket) is
$$
\LL_UV^M=U^N\*_NV^M-(\d^M_Q\d^P_N-\eta^{MP}\eta_{NQ})\*_PU^QV^N
=L_UV^M+\eta^{MP}\eta_{NQ}\*_PU^QV^N
\komma\Eqn\DoubledDiffeomorphism
$$
where $\eta$ is the $O(d,d)$-invariant metric.
Comparing to the different forms for the expression in the U-duality
setting, we see that it has analogous properties --- the expression
$\d^M_Q\d^P_N-\eta^{MP}\eta_{NQ}$ projects on the adjoint, and the
section condition is simply $\eta^{MN}\*_M\otimes\*_N=0$. The section
condition is fulfilled by any pairs of covectors in an isotropic
(null) subspace of dimension $d$.

\subsection\GenDiffAlgebra{The algebra of 
generalised diffeomorphisms}The tensor $Y$ in the Ansatz for the
generalised diffeomorphisms is completely determined by demanding that
the transformations form an algebra. We need
$$
[\LL_U,\LL_V]=\LL_{\leftbr U,V\rightbr}\komma\Eqn\TheAlgebra
$$
where 
$$
\leftbr
U,V\rightbr=[U,V]^M+\fr2Y^{MN}{}_{PQ}(\*_NU^PV^Q-\*_NV^PU^Q)\punkt\eqn
$$
This bracket is analogous to and shares many properties with the
Courant bracket in doubled geometry, 
and may be called an ``exceptional Courant bracket''.
A direct calculation shows that the algebra closes according to
eq. (\TheAlgebra) iff the following conditions are fulfilled:
\multi
$$
\eqalignno{
&Y^{MN}{}_{PQ}\*_M\otimes\*_N=0\komma&\multiEqn\ConditionsOnYa{a}\cr
&\left(Y^{MN}{}_{TQ}Y^{TP}{}_{RS}-Y^{MN}{}_{RS}\delta^P_Q\right)
\*_{(N}\otimes\*_{P)}=0\komma&\multiEqn\ConditionsOnYb{b}\cr
&\left(Y^{MN}{}_{TQ}Y^{TP}{}_{[SR]}+2Y^{MN}{}_{[R|T|}Y^{TP}{}_{S]Q}\right.\cr
&\qquad\left.-Y^{MN}{}_{[RS]}\delta^P_Q-2Y^{MN}{}_{[S|Q|}\delta^P_{R]}\right)
\*_{(N}\otimes\*_{P)}=0\komma&\multiEqn\ConditionsOnYc{c}\cr
&\left(Y^{MN}{}_{TQ}Y^{TP}{}_{(SR)}+2Y^{MN}{}_{(R|T|}Y^{TP}{}_{S)Q}\right.\cr
&\qquad\left.-Y^{MN}{}_{(RS)}\delta^P_Q-2Y^{MN}{}_{(S|Q|}\delta^P_{R)}\right)
\*_{[N}\otimes\*_{P]}=0\punkt&\multiEqn\ConditionsOnYd{d}\cr
}\multiEqnAll\ConditionsOnY
$$
Eq. (\ConditionsOnYa), which roughly speaking is the section condition of
the previous subsection, 
comes from a single remaining term in the
commutator containing a derivative. Equation (\ConditionsOnYb) 
comes from terms with
$\*^2UV$ and $U\*^2V$, while eqs. (\ConditionsOnYc) 
and (\ConditionsOnYd) multiply $\*U\*V$ (hence the
opposite symmetrisations). No specific symmetry properties have been
assumed for $Y$. 

The easiest possible expression for $Y^{MN}{}_{PQ}$, satisfying the section
condition, would be that it is proportional to the projector on $R_2$ itself,
$$
Y^{MN}{}_{PQ}=kP_{(R_2)}^{MN}{}_{PQ}\punkt\Eqn\YUpToSix
$$ 
This is indeed the case for $n\leq6$, where it then immediately follows
that $Y$ is symmetric in pairs of indices. For higher $n$, $Y$ also
contains some other term that separately vanishes with the help of the
section condition when contracted with derivatives as in
eq. (\ConditionsOnYa).

In all cases up to $n=6$,
with $Y$ given by eq. (\YUpToSix), the equations (\ConditionsOnY)
simplify. Terms with coefficients 1 and 2 in the third and fourth
equation combine into symmetrisations in three
indices. Note that this is also precisely what is needed for the
second equation to hold; if
$$
Y^{(MN}{}_{TQ}Y^{P)T}{}_{RS}-Y^{(MN}{}_{RS}\delta^{P)}_Q=0\komma\Eqn\SimplestRel
$$
which turns out to be true for $n\leq5$, but needs a little
modification for $n=6,7$, the indices on the derivative
can be cycled to $Y$, and the equations are then satisfied thanks to
the section condition.

Let $Y^{MN}{}_{PQ}=kP_{(R_2)}^{MN}{}_{PQ}$,
where $P_{(R_2)}$ is
the projector on $R_2$ in the symmetric product of two $R_1$'s (this
will be modified for $n=7$). Eq. (\SimplestRel) has the structure
$R_1\otimes R_2\otimes(\otimes_s^3\bar R_1)$. The number of possible
invariant tensors is the number of singlets in this tensor product,
\ie, the number of irreducible modules in 
$T=(R_1\otimes R_2)\cap(\otimes_s^3R_1)$. 

For $n\leq5$, this number is 1,
which means that the two terms in eq. (\SimplestRel) are
proportional to each other. Then the constant $k$ can be determined
simply by taking some trace of the equation. One gets $k=2(n-1)$. The
explicit expressions are given below.

For $n=6,7$, the number of irreducible
modules in $T$ is $2$. 
In $E_6$, there is an invariant symmetric tensor $c^{MNP}$. If one
normalises it so that $c^{MNP}c_{MNP}=27$ (in which case
$P_{({\bf27})}^{MN}{}_{PQ}=c^{MNR}c_{PQR}$), the relevant identity reads
$$
10P_{({\bf27})}^{(MN}{}_{QT}P_{({\bf27})}^{P)T}{}_{RS}
-P_{({\bf27})}^{(MN}{}_{RS}\delta^{P)}_Q
-\fr3c^{MNP}c_{QRS}=0\punkt\Eqn\ESixEqn
$$
The last term is of course $\bf27$- or $\overline{\bf27}$-projected on
any pair of indices. The 
tensor $Y$ is given by the same expression as for lower $n$.

When $n=7$, $R_1$ is the fundamental 56-dimensional module. It is
symplectic, so there is an invariant tensor $\e_{MN}$. We choose
conventions where $\e^{MN}$ is the inverse to $\e_{MN}$. $R_2$ is $\bf
133$, the adjoint. There is a completely symmetric 4-index tensor
$c^{MNPQ}$, which we choose to normalise so that $c^{MNPQ}=P_{({\bf133})}^{(MNPQ)}$.
Then, the projector on $\bf133$ can be written
$$
P_{({\bf133})}^{MN}{}_{PQ}=c^{MN}{}_{PQ}+\fr{12}\d^{(M}_P\d^{N)}_Q\komma\eqn
$$  
which is a practical expression when one wants to move indices on $P$.
The relevant identity generalising eqs. (\SimplestRel,\ESixEqn) is
$$
12P_{({\bf133})}^{(MN}{}_{QT}P_{({\bf133})}^{P)T}{}_{RS}
-4c^{MNPT}P_{({\bf133})TQRS}
-P_{({\bf133})}^{(MN}{}_{RS}\d^{P)}_Q=0\punkt\eqn
$$
The tensor $Y$ now necessarily contains an antisymmetric part, and
takes the form
$$
Y^{MN}{}_{PQ}=12P_{({\bf133})}^{MN}{}_{PQ}+\fr2\e^{MN}\e_{PQ}\punkt\eqn
$$
It is clear (even more so from the argument in Section 4.5) 
that if an $SL(7)$ vector is picked out by the section
condition, one will also have $\e^{MN}\*_M\otimes\*_N=0$. 

In Section 2.4, we will study the case $n=8$, 
for which the generalised Lie derivatives fail to form an ordinary Lie algebra.

To summarise, the forms of the tensor $Y$ in the different cases are:
$$
\matrix{
n=3:\hfill&Y^{i\a,j\b}{}_{k\g,l\d}=4\d^{ij}_{kl}\d^{\a\b}_{\g\d}\komma\hfill\cr
n=4:\hfill\qquad&Y^{mn,pq}{}_{rs,tu}=
            6\d^{mnpq}_{rstu}\komma\hfill\cr
n=5:\hfill&Y^{\a\b}{}_{\g\d}=\fr2\g_a^{\a\b}\g^a_{\g\d}
\komma\hfill\cr
n=6:\hfill&Y^{MN}{}_{PQ}=10c^{MNR}c_{PQR}\komma\hfill\cr
n=7:\hfill&Y^{MN}{}_{PQ}=12c^{MN}{}_{PQ}+\d^{(M}_P\d^{N)}_Q+\fr2\e^{MN}\e_{PQ}
\komma\hfill\cr}
\eqn
$$
with index notation that is hopefully self-explanatory.

In all cases, it can be
checked that if one makes a Fierz-like rearrangement and 
rewrites $Y^{MN}{}_{PQ}$ in a basis 
where $R_1\otimes\bar R_1$ represented by the indices $N$ and $P$
(which are the ones contracting $\*_NU^P$) is
expanded in irreducible modules, one gets
$$
Y^{MN}{}_{PQ}=-\a_nP_{(adj)}{}^M{}_{Q,}{}^N{}_P+\b_n\d^M_Q\d^N_P
+\d^M_P\d^N_Q\punkt\eqn
$$
Here, the last term cancels the second term in the ordinary Lie
derivative, when inserted in eq. (\LieDerAnsatz). The projector on the
adjoint is defined so that 
$P_{(adj)}{}^M{}_{N,}{}^R{}_SP_{(adj)}{}^S{}_{R,}{}^P{}_Q
=P_{(adj)}{}^M{}_{N,}{}^P{}_Q$
and $P_{(adj)}{}^M{}_{N,}{}^N{}_M=\hbox{dim}(adj)$
(unfortunately, the convention for raising and lowering indices 
leads to $P_{(adj)}=-P_{\bf133}$ for $n=7$).
The constants $\a_n$ and $\b_n$ take the numerical values
$(\a_4,\b_4)=(3,\fr5)$, $(\a_5,\b_5)=(4,\fr4)$, 
$(\a_6,\b_6)=(6,\fr3)$,
$(\a_7,\b_7)=(12,\fr2)$. 
For $n=3$, the U-duality group $SL(3)\times SL(2)$ is not
semisimple. There one has
$$
Y^{i\a,j\b}{}_{k\g,l\d}=\d^i_k\d^j_l\d^\a_\g\d^\b_\d
-(2P_{\bf(8,1)}+3P_{\bf(1,3)})^{i\a}{}_{l\d,}{}^{j\b}{}_{k\g}
+\fr6\d^i_l\d^j_k\d^\a_\d\d^\b_\g\punkt\eqn
$$
This provides the precise relations with the
expressions of ref. [\CoimbraStricklandWaldram]. There seems to be a pattern:
$\b_n=\fr{9-n}$. The coefficient in front of $P_{(R_2)}$ is always $2(n-1)$.

\subsection\Jacobiator{The Jacobi identity}For an ordinary Lie
derivative, $L_UV$ is  
already antisymmetric in $U$ and $V$, so $[U,V]=L_UV$. This is now no
longer the case. Define the symmetric part of $\LL_UV$ as
$\leftpar U,V\rightpar=\fr2(\LL_UV+\LL_VU)$. Then
$$
\LL_{\leftpar
U,V\rightpar}W^M=-\left(
Y^{M[N}{}_{PQ}Y^{|P|R]}{}_{[ST]}+Y^{M[N}{}_{[ST]}\d^{R]}_Q\right)
\*_NU^S\*_RV^TW^Q\komma\eqn
$$
after eq. (\ConditionsOnYb) has been used for the
part symmetric in indices on derivatives.
This vanishes trivially for $n\leq6$ and equals
$-\fr4\e^{NR}\e_{PQ}\*_{[N}U^P\*_{R]}V^QW^M=0$ for $n=7$ by the
section condition.
Hence a generalised diffeomorphism generated by $\leftpar
U,V\rightpar$ gives a zero 
transformation. 

The Jacobiator $\leftbr\cdot,\cdot,\cdot\rightbr$ 
can be calculated using the same method as in ref. [\HitchinLectures].
Let 
$\leftbr U,V,W\rightbr=\leftbr\leftbr U,V\rightbr,W\rightbr
+\hbox{cycl.}$. Using
$$
\eqalign{
\leftbr\leftbr U,V\rightbr,W\rightbr
&=\fr2(\LL_{\leftbr U,V\rightbr}W-\LL_W\leftbr U,V\rightbr)\cr
&=\fr2(\LL_U\LL_VW-\LL_V\LL_UW)-\fr4(\LL_W\LL_UV-\LL_W\LL_VU)\komma\cr
}\Eqn\JacobiatorHelpEq
$$
we see that the Jacobiator can be written using either the first or
the second term of the first line of eq. (\JacobiatorHelpEq):
$$
\leftbr U,V,W\rightbr=\left\{\matrix{
\fr4\LL_{\leftbr U,V\rightbr}W+\hbox{cycl.}\hfill\cr
\fr2\LL_W\leftbr U,V\rightbr+\hbox{cycl.}\hfill\cr
}\right.
\eqn
$$
and thus also as
$$
\leftbr U,V,W\rightbr
=\fr6(\LL_{\leftbr U,V\rightbr}W+\LL_W\leftbr
U,V\rightbr)+\hbox{cycl.}
=\fr3\leftpar\leftbr U,V\rightbr,W\rightpar
+\hbox{cycl.}\punkt\eqn
$$
So, the Jacobiator generates a zero transformation for all
$n\leq7$. We will give a more careful interpretation of this statement
in the following section.

\subsection\NisEight{$n=8$}For $n=8$, the algebra does not work. This
is more or less expected, since this is where the dual (in the
11-dimensional sense) gravity field
becomes relevant. As we will see in the following section, the
counting of gauge parameters nevertheless matches the ones in component
form, including the ``dual diffeomorphisms''. For this reason, we
would still like to say a few words about the failure of the algebra.

The section condition will necessarily be in $\bar
R_2={\bf1}\oplus{\bf3875}$, as can be seen at an early stage in the
calculation (this is vindicated by the entry in Table
\ReducibilityTable, as it will be calculated in Section 3). 
This leaves only ${\bf27000}$ in the symmetric
product of two derivatives, and it can be deduced that it also
implies that the projections of two derivatives on $\bf248$
vanish. The projection operators are [\KoepsellNicolaiSamtleben]
$$
\eqalign{
&P_{({\bf1})}^{MN}{}_{PQ}=\fr{248}\eta^{MN}\eta_{PQ}\komma\cr
&P_{({\bf248})}^{MN}{}_{PQ}=-\fr{60}f_A{}^{MN}f^A{}_{PQ}\komma\cr
&P_{({\bf3875})}^{MN}{}_{PQ}=-\fr{14}f^{A(M}{}_Pf_A{}^{N)}{}_Q
    +\fr7\d^{(M}_P\d^{N)}_Q-\fr{56}\eta^{MN}\eta_{PQ}\komma\cr
}\eqn
$$
where the structure constants are normalised so that
$f^{MAB}f_{NAB}=-60\d^M_N$. 

There is a certain combination of these
projectors that combine in the same way as for lower $n$, and this is
a reasonable candidate: 
$$
\eqalign{
\LL_UV^M
&=L_UV^M+(14P_{(\bf3875)}-30P_{(\bf248)}+62P_{(\bf1)})^{MN}{}_{PQ}\*_NU^PV^Q\cr
&=U^N\*_NV^M+\*_NU^NV^M-f^{AM}{}_Qf_A{}^N{}_P\*_NU^PV^Q\punkt\cr
}\Eqn\EEightAnsatz
$$
Note that the coefficients follow the pattern deduced from lower
$n$. 
It certainly looks a lot easier to try this than to make a general
Ansatz with the $P$'s. The section condition now implies
$$
\eqalign{
&\eta^{MN}\*_M\otimes\*_N=0\komma\cr
&f^{AMN}\*_M\otimes\*_N=0\komma\cr
&(f^{AM}{}_Pf_A{}^N{}_Q-2\d^{(M}_P\d^{N)}_Q)\*_M\otimes\*_N=0\cr
}\eqn
$$
(note that symmetrisation is not needed in the first term in the last
equation).

The terms in $([\LL_U,\LL_V]-\LL_{\leftbr
U,V\rightbr})W$ with a derivative on $W$ vanish due to the section
condition. It turns out that also the terms of the form ``$\*U\*VW$''
all cancel. A long calculation leads to a single remaining obstruction
with the structure
$f^{MN}{}_Qf^P{}_{ST}\*_N\*_PU^SV^TW^Q-(U\leftrightarrow V)$.
Even if our guess (\EEightAnsatz) was dictated by the tensor-friendly form
(\AlgebraTransf), it can be checked explicitly that other combinations
fail even more seriously to fulfill an algebra.

\vfill\eject
\section\Reducibility{The ghost tower}The tensor gauge transformations
are reducible. A 2-form 
transformation has a 1-form reducibility and a 0-form second order
reducibility, so that the effective number of gauge parameters in $n$
dimensions is
${n\choose 2}-n+1={n-1\choose2}$, and analogously for a a 5-form
parameter ${n-1\choose5}$. Including diffeomorphisms, the effective number of 
generalised diffeomorphisms should be $n+{n-1\choose2}+{n-1\choose5}$,
as long as dual gravity does not enter. This number will be checked by
examining the reducibility of the generalised diffeomorphisms in their
covariant form. 

In doubled geometry, the gauge transformations contained in the
generalised diffeomorphisms are diffeomorphisms and the gauge
transformation of the 2-form $B$, $\delta B=d\Lambda$. The latter is
reducible, since $\Lambda=d\phi$ gives rise to no transformation on
the fields. This reducibility is directly reflected in the
reducibility of the generalised diffeomorphisms: a parameter
$U^M=\eta^{MN}\*_N\xi$ does not enter the transformation
(\DoubledDiffeomorphism) due to the section condition.  

Analogously, the reducibility in the exceptional setting is also associated
with the section condition. A parameter constructed as 
$U^M[\xi]=\*_N\xi^{MN}$, where $\xi$ is in the
representation $R_2$ conjugate to the section condition, will generate a zero
transformation through $\LL_{U[\xi]}$. 
This is easily seen from the form (\AlgebraTransf) of the generalised
diffeomorphisms. In the transport term, $\*_P\xi^{NP}\*_P=0$ due to
the section condition, and in the $E_{n(n)}\times\RR$ transformation
terms, $\*^2\xi$ contains neither the singlet nor the adjoint.
This is the first order
reducibility. The relation for $U[\xi]$ will in turn be reducible, in
the sense that for an $\xi^{MN}=\*_P\xi'^{MNP}$, with $\xi'$ in a certain
representation $R_3$, $U[\xi[\xi']]=0$, and so on. 

We will now examine the representation content of the reducibility,
and show that it is directly connected to the properties of the 
weak section condition.
Namely, consider an object $\l_M$ in $\bar R_1$, satisfying the weak
section condition, $T\equiv\l^2|_{\bar R_2}=0$.
This constraint is reducible, there is always some representation
$\bar R_3$ such that $T'\equiv(\l T)|_{\bar R_3}=0$. Again, given this
form of $T'$ it will satisfy $T''\equiv(\l T')|_{\bar R_4}=0$, and so
on. The representations in question can be determined by examining the
partition function for the object $\l$, as will be done below. A
typical example is provided by pure spinors in $D=10$, which will
actually be one of the cases.

Now, consider a momentum in $R_1$ conjugate to $\l$, and call it
$w^M={\*\over\*\l_M}$. A naked $w$ is not invariant with respect to the
constraint on $\l$, $\l^2|_{\bar R_2}=0$. This constraint, with
parameter $\xi^{MN}$ in $R_2$, generates a transformation of $w$,
$\d_\xi w^M=[-\fr2\xi^{NP}\l_N\l_P,w^M]=\l_N\xi^{MN}$. Here we note
that this transformation is formally equivalent to the reducibility of
the parameter of generalised diffeomorphisms, if we replace $\l_M$ by
$\*_M$ and $w^M$ by $U^M$. Once this is established, it is clear that
the parameters of higher order reducibility
are identical to those of the reducibility of the weak section condition on
$\*$.
The first order reducibility is
precisely such that it leaves both the scalar and adjoint parts of
$\*_MU^N$ invariant. 
This point of view gives yet another, algebraic, reason why only the scalar and
adjoint of $\*U$ can appear in the generalised Lie derivative (\AlgebraTransf).

Such towers of ghosts have
been examined in the cases of pure spinors in various dimensions. The
details can be derived as follows.
Write a partition function by counting the
homogeneous functions of degree $i$ of the constrained object $\l$:
$$
Z(t)=\sum\limits_{i=0}^\infty\hbox{dim}(r_i)t^i\punkt\Eqn\PartitionDef
$$
(This, and everything below, can of course in principle 
be refined by not only counting dimensions, but
also actual representations. For low $R_k$'s, the actual
representations can however be deduced safely by just observing the dimension.)
In all the cases under consideration, the weak section condition is such that
the representation $r_i$ contained in the $i$'th power of $\l$ 
is the irreducible representation with
highest weight $i$ times the one of the coordinate representation
$\bar R_1$. This is a direct consequence of the fact that all smaller
representations than the largest one are absent in $r_2$ due to
the bilinear constraint. In $r_i$, any smaller representation would
have to be formed by tensor some pair of $\l$'s into such a smaller
representation.  

These partition functions $Z_n$ are given for different values of $n$ below. 
They can either be calculated with the help of explicit 
expressions\foot\dagger{See for example The Online Encyclopedia of
Integer Sequences, {\xtt http://oeis.org}.}\ 
for $\dim(r_i)$, or alternatively in a
pragmatic way: by forming the series in eq. (\PartitionDef) from
dimensions calculated with LiE, up to some high power where we
safely can conclude that it coincides with 
$(1-t)^{-(\hbox{\xrm some number})}\times F(t)$, where $F(t)$ is a
polynomial with $F(1)\neq0$.
\multi
$$
\eqalignno{
Z_3(t)&=\sum\limits_{i=0}^\infty(i+1){i+2\choose2}t^i=(1-t)^{-4}(1+2t)\komma\cr
Z_4(t)&=\sum\limits_{i=0}^\infty
  {1\over3}{i+4\choose4}{i+3\choose2}t^i=(1-t)^{-7}(1+3t+t^2)\komma\cr
Z_5(t)&=\sum\limits_{i=0}^\infty
  {1\over10}{i+7\choose7}{i+5\choose3}t^i=(1-t)^{-11}(1+t)(1+4t+t^2)\komma\cr
Z_6(t)&=\sum\limits_{i=0}^\infty
   {1\over56}{i+11\choose11}{i+8\choose5}t^i
=(1-t)^{-17}(1+t)(1+9t+19t^2+9t^3+t^4)\komma
                    &\multieqn{}\cr
Z_7(t)&=\sum\limits_{i=0}^\infty
{1\over3^2\cd5\cd11\cd13} 
  (i+9){i+17\choose17}{i+13\choose9} t^i\cr
&=(1-t)^{-28}(1 + 28 t + 273 t^2 + 1248 t^3 + 
3003 t^4 + 4004 t^5 + 3003 t^6 + 
 1248 t^7 \cr
&\qquad+ 273 t^8 + 28 t^9 + t^{10})\komma\cr
Z_8(t)&=\sum\limits_{i=0}^\infty
{1\over2\cd7\cd11^2\cd13\cd17\cd19^2\cd23\cd29}(2i+29)
{i+28\choose28}{i+23\choose18}{i+19\choose10}t^i\cr
&=(1-t)^{-58}(1+t)(1 + 189 t + 14\,080 t^2 + 562\,133 t^3 + 13\,722\,599 t^4 \cr
&\qquad+
220\,731\,150 t^5 +  
 2\,454\,952\,400 t^6 + 19\,517\,762\,786 t^7 + 113\,608\,689\,871 t^8 \cr
&\qquad+ 
 492\,718\,282\,457 t^9 + 1\,612\,836\,871\,168 t^{10} +
 4\,022\,154\,098\,447 t^{11} \cr 
&\qquad+ 
 7\,692\,605\,013\,883 t^{12} + 11\,332\,578\,013\,712 t^{13} +
 12\,891\,341\,012\,848 t^{14} \cr 
&\qquad+ 
 11\,332\,578\,013\,712 t^{15} + 7\,692\,605\,013\,883 t^{16} +
 4\,022\,154\,098\,447 t^{17} \cr 
&\qquad+ 
 1\,612\,836\,871\,168 t^{18} + 492\,718\,282\,457 t^{19} +
 113\,608\,689\,871 t^{20} \cr 
&\qquad+ 
 19\,517\,762\,786 t^{21} + 2\,454\,952\,400 t^{22} + 220\,731\,150 t^{23} + 
 13\,722\,599 t^{24} \cr
&\qquad+ 562\,133 t^{25} + 14\,080 t^{26} + 189 t^{27} + t^{28})\punkt\cr
}
$$
The effective number of independent gauge parameters is read off as
the negative power of the first factor (the number of ``bosonic
degrees of freedom''). For $n\leq6$ the corresponding spaces and 
their dimensions are known earlier. 
For $n=4$, the 7-dimensional space is a c\^one over the Grassmannian
$Gr(2,5)$ of 2-planes in 5 dimensions. 
For $n=5$, 11 is the dimension of the space of pure spinors of
$Spin(5,5)$. For $n=6$, an object $X^M$ with $c_{MNP}X^NX^P=0$ lies on a
17-dimensional c\^one over the 16-dimensional 
Cayley plane [\CederwallJordanMech].

For $n\leq7$, the dimension is 1 greater than the dimension of $R_1$
for the next lower value of $n$. This observation should be related to
the existence of a 3-grading of the algebra corresponding to the
subgroup $E_{n+1(n+1)}\supset E_{n(n)}\times{\Bbb R}$, providing a non-linear
``conformal'' realisation of $E_{n+1(n+1)}$ on $R_1$ of $E_{n(n)}$
[\GunaydinConformal,\GunaydinKoepsellNicolai].
In fact, the present construction provides an infinite-dimensional
linear representation of 
$E_{n+1(n+1)}$ on polynomials of the constrained objects in $R_1$ of
$E_{n(n)}$, \ie, on $\oplus_{i=0}^\infty(i0\ldots0)$
(the Dynkin index for $R_1$ is taken to be $(10\ldots0)$), which can be
thought of as a singleton representation. For $n=5$, this was also
observed in ref. [\PiolineWaldron].
For $n=7$, the
grading corresponding to $E_{8(8)}\supset E_{7(7)}\times SL(2,{\Bbb
R})$ is a 5-grading [\GunaydinKoepsellNicolai].
The dimensions can also be identified as the dimensions of coadjoint
nilpotent orbits of $\fr2$-BPS instantons [\GreenMillerVanhove].

For $n\leq7$, the number of gauge parameters thus calculated 
match the number of diffeomorphisms,
2-form and 5-form (for $n\geq6$) transformations calculated above. For
$n=8$, strikingly enough, the counting also matches if one includes
also $n{n-1\choose7}=8$ gauge parameters for the vector-valued 7-form
transformations of the dual gravity field. The counting, and the
comparison with the results from reducibility are summarised in
Table 2. 

\vskip4\parskip
\ruledtable
$n$\|diffeo|2-form|5-form|dual diffeo|total\crthick
3\|3|1|||4\cr
4\|4|3|||7\cr
5\|5|6|0||11\cr
6\|6|10|1||17\cr
7\|7|15|6|0|28\cr
8\|8|21|21|8|\hskip4pt 58
\endruledtable
\Table\ParameterCountTable{The counting of gauge parameters.}

More information on
the structure of the reducibility can be extracted by rewriting the
partition functions as products of ghost
partitions,
$$
Z(t)=\prod_{k=1}^\infty(1-t^k)^{-A_k}\komma\eqn
$$
where the power $A_k$ is $(-1)^{k-1}$ times the dimension of the $(k-1)$'th
reducibility ghost representation. To get the number of effective
gauge transformations, we want to calculate a regulated
sum $\sum_{k=1}^\infty A_k$. Taking the logarithm,
$$
\log Z(t)=-\sum_{k=1}^\infty A_k\log(1-t^k)
=-\sum_{k=1}^\infty A_k\left(\log(1-t)+\log\sum_{i=0}^{k-1}t^i\right)
\punkt\eqn
$$
The second logarithm is regular at $t=1$, so the sum is obtained as
the coefficient of the singular behaviour $-\log(1-t)$ at $t=1$, as
argued above.
This result is also what one obtains from regulating the sum with
analytic continuation.
A more refined treatment (see for example
ref. [\BerkovitsNekrasovCharacter]) is required if one wants to
calculate other moments like ghost number $\sum_{k=1}^\infty kA_k$. 

A completely refined partition function gives information about
the exact representations $R_k$. It requires rewriting the known
partition function, including complete information about representations,
$$
{\cal Z}(t)=\bigoplus_{i=0}^\infty (i0\ldots0)t^i\punkt\eqn
$$
on a product form
$$
{\cal Z}(t)=\left(\prod_{k\in2{\Bbb N}+1}\bigoplus_{j=0}^\infty t^{jk}\otimes_s^j
R_k\right)
\otimes\left(\prod_{k\in2{\Bbb N}+2}\bigoplus_{j=0}^{\dim(R_k)} 
(-1)^jt^{jk} \wedge^j
R_k\right)\komma\eqn
$$
where the first factor contains partitions for bosons in $R_{odd}$
and the second one partitions for fermions in $R_{even}$. 
This can be done recursively to find arbitrary $R_k$.

Unlike tensor gauge transformations and generalised diffeomorphisms in
doubled field theory, the tower of ghosts
(reducibility) is infinite in all cases. Such a statement can of
course change if one is allowed to break $E_n$ invariance to some
smaller covariance. Note that the representations $R_k$, listed in
Table \ReducibilityTable,
coincide with the representations of ``form fields'', listed in various
tables (see \eg\
refs. [\deWitNicolaiSamtleben,\deWitSamtleben,\PalmkvistHierarchy,\BermanMusaevThompson])\foot\star{The
representations we derive coincide exactly with those appearing in
Borcherds algebras. For $n=8$, we have verified this up to $R_4$. The
reason for this is probably that Serre relations for the Borcherds
algebra is effectively encoded in an algebraic constraint (the section
condition). We may come back to this in a future publication.}. 
The representations $R_k$ are possible representations for
$k$-form fields in the uncompactified $11-n$ dimensions. The sequences
continues beyond those of the form fields, and do not halt at any
finite $k$.

For example, when $n=6$,
$$
Z_6(t)=(1 - t)^{-27} (1 - t^2)^{27} (1 - t^3)^{-78} (1 - t^4)^{351} (1 - 
    t^5)^{-1755} (1 - t^6)^{8983} (1 - t^7)^{-47034} 
\times\ldots\eqn
$$
Here, we recognise the $\bf27$, $\overline{\bf27}$, $\bf78$,
$\overline{\bf351'}$ and 
$\overline{\bf1728}\oplus\overline{\bf27}$ from the table of
fields\foot\dagger{We write ${\bf351'}$ for a tensor
$A^{[MN]}$. There are four 351-dimensional representations of $E_6$: this one,
the symmetric ${\bf351}$ of a tensor
$S^{(MN)}$ with $c_{MNP}S^{NP}=0$, and their conjugates.}.

The Dynkin indices of $R_k$ for the first few $k$ are depicted in
Figure 1. For $n=7,8$, this gives the leading (biggest)
representations. For $n=3$, $R_1=(10)(1)$ (\ie, 1's at the nodes
marked $R_4$ and $R_3$ in the figure), but $R_{2,3,4}$ are given
accurately by the figure.

\vskip3\parskip
\vtop{\epsffile{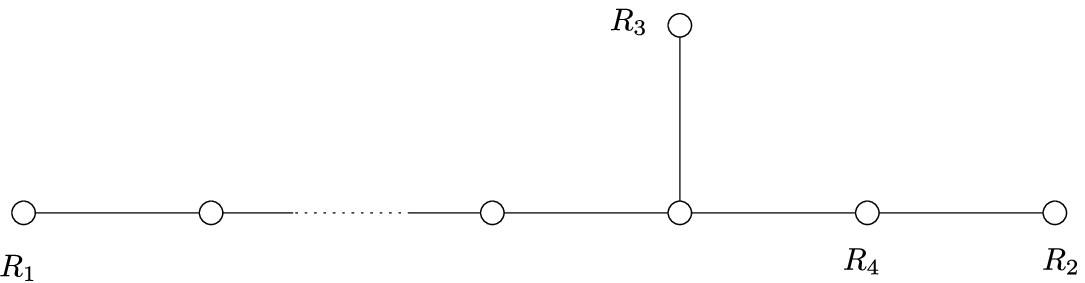}\hfill\break
\centerline{\it Figure 1: Dynkin indices of some reducibility representations.}
}
\vskip2\parskip

Let us spell out an example in some detail. For $n=5$, the parameter
is a spinor, and we have  
$$
\eqalign{
Y^{\a\b}{}_{\g\d}&=8P_{\bf10}^{\a\b}{}_{\g\d}
=\fr2\g_a^{\a\b}\g^a_{\g\d}\cr&=\d^\a_\g\d^\b_\d
  +\fr8(\g^{ab})^\a{}_\d(\g_{ab})^\b{}_\g+\fr4\d^\a_\d\d^\b_\g\komma\cr
}\eqn
$$
so that generalised diffeomorphisms are generated by $\LL_U$, where
$$
\LL_UV^\a=(U\*)V^\a+\fr8(\*\g^{ab}U)(\g_{ab}V)^\a+\fr4(\*U)V^\a\punkt\eqn
$$
Now, consider a parameter $U$ which is constructed as a derivative of
a $Spin(5,5)$ vector as 
$$
U^\a[\xi]=\g_a^{\a\b}\*_\b\xi^a\punkt\Eqn\ExplRed
$$
Substituting
this parameter in the transformation gives
$$
\LL_{U[\xi]}V^\a=\g_a^{\b\g}\*_\g\xi^a\*_\b V^\a
+\fr8(\*\g^{ab}\g_c\*)\xi^c(\g_{ab}V)^\a+\fr4(\*\g_c\*)\xi^cV^\a\punkt\eqn
$$
All three terms vanish, since the section condition implies
$\g_a^{\a\b}\*_\a\otimes\*_\b=0$. 
So, the reducibility of the parameter $U^\a$ lies at least in
$\xi^a$ (it is not difficult to show that this the complete
reducibility). 
Then $\xi$ in turn has a second order reducibility,
$\xi^a=(\*\g^a\xi')$, which gives zero in eq. (\ExplRed) by virtue of
the Fierz identity $\g^{a\a(\b}\g_a{}^{\g\d)}=0$ and the section
condition. Next,
$\xi'_\a=(\g_{ab}\*)_\a\xi''{}^{ab}$, etc. The structure is reflected
in the product form
$$
Z_5(t)=(1 - t)^{-16} (1 - t^2)^{10} (1 - t^3)^{-16} (1 - t^4)^{45} (1 - 
    t^5)^{-144} (1 - t^6)^{456} (1 - t^7)^{-1440}\times\ldots\eqn
$$

The parameter $U$ in $R_1=\bf16$ 
branches into the diffeomorphism vector and a 2-form
and a 5-form gauge transformation.  
The vector $\xi$ in $R_2=\bf10$ branches into a 1-form and a 4-form when
$Spin(5,5)\rightarrow GL(5)$. This is of course the
reducibility one wants for a 2-form and a 5-form gauge
transformation. 
At the next level,
something has to change, however. The second order reducibility in the
$GL(n)$ language contains a scalar and a 3-form (in total 11 second
order ghosts), while the smallest representation of $\xi'$ such that
$\xi=\*\xi'$ is $R_3=\bf\overline{16}$. There is an excess of a
1-form. This does not imply any extra reducibility as compared to the
$GL(n)$-covariant considerations, but has to be compensated for by higher
reducibilities. The reducibility becomes infinite, with ever growing
representations $R_k$, but in a way that makes the resulting infinite
alternating sum meaningful.

\section\SectionCondition{The section condition as a linear
constraint}In the generalised space-time the coordinates 
form the representation
 $R_1$ of the duality group $E_{n(n)}$.  The section condition is given by
 projecting out a particular representation $\bar R_2$: 
 $$
 (\partial  \otimes \partial) |_{\bar R_2} = 0 \punkt 
 \eqn
$$ 
 We propose instead to introduce an auxiliary object $\Lambda$ in a
 representation $T$ which will play the analogous role to the pure
 spinor in the $O(d,d)$ case and pick out a subspace akin to the way
 the pure spinor identifies a maximal isotropic subspace.  The ``pure spinor''
 constraints in this case transform in a representation $P$, \ie, has
 the form 
$$
\Lambda^2 |_{P}= 0 \punkt
\eqn
$$ 
 The linear section equation will have the form 
 $$
 (\Lambda \otimes \partial) |_N  = 0  
\Eqn\LinearSectionCondition
 $$
 for some representation $N$, and will imply the section
 condition.  

We will make this reformulation for $n\leq7$. In all cases, the
representation $T$ of
$\Lambda$ is $\bar R_3$, and the representation $N$ for the linear constraint
of eq. (\LinearSectionCondition) is $\bar R_4$. The constraint on $\Lambda$
is absent for $n\leq4$. We are not yet able to deduce a pattern for
the representation $P$ of this constraint.
 
 Let us remind of the situation in doubled generalised geometry. The
 section condition in its non-linear form reads
 $\eta^{MN}\*_M\otimes\*_N=0$. The largest linear subspace of the
 c\^one of null vectors is an isotropic subspace. Such a space is
 determined by the choice of a pure $Spin(d,d)$ spinor $\L^\a$, obeying 
$(\L\g^M\L)=0$, and the strong section condition is replaced by the
 linear condition $(\g^M\L)_\a\*_M=0$.
 
 \subsection\NNisThree{$n=3$}The duality group is $G= SL(3) \times
 SL(2)$.  The six coordinates of extended space time are in the
 $R_1=({\bf3},{\bf2})$ representation of this duality group.  
The section condition reads
 $$
  \e^{\a \b} \partial_{a\a} A \partial_{b\b} B = 0 \komma \quad \a=1,2 \quad
 a = 1,2,3 \punkt \Eqn\WSCnThree
 $$
 Let us instead introduce a $\L^\a$ in $\bar R_3={\bf(1,2)}$ 
and consider the linear section equation
 $$
 \L^\a \partial_{a\a} = 0 \punkt\Eqn\LSCnThree
 $$
 This transforms in $\bar R_4=(\overline{\bf3},{\bf1})$ and it implies
 the section 
 condition. For instance, choose a frame where 
$\L^1 \neq 0$ with $\L^2=0$, then we see
 that eq. (\LSCnThree) would imply that $\partial_{a1} = 0 $ and clearly
 then eq. (\WSCnThree) is satisfied.  
The linear section condition serves to reduce from the six extended
 coordinates to three physical coordinates.

 \subsection\NNisFour{$n=4$}In this case the duality group is $G=
 SL(5)$.  
The ten coordinates of extended space time are in the $R_1=\overline{\bf10}$
representation of
$G$.  The section condition is in  $\bar R_2={\bf5} $:  
 $$
 \e^{a b c d e} \partial_{ab} A \partial_{cd} B = 0 \komma  \quad a =
 1\dots 5 \punkt \Eqn\WSCnFour
 $$
 Instead let us introduce an $\L_a$ in the $\bar R_3=\overline{\bf 5}$ 
and consider the linear section equation
 $$
 \L_{[a} \partial_{bc]} = 0 \punkt\Eqn\LSCnFour
 $$
 This transforms in  $\bar R_4={\bf10}$ and it implies the
 section condition.  For instance choose $\L_1 \neq 0$ with other
 components vanishing. Then we see that eq. (\LSCnFour) would imply
 that
 $\partial_{23}=\partial_{24}=\partial_{25}=\partial_{34}
 =\partial_{35}=\partial_{45} 
 = 0 $ and clearly then eq. (\WSCnFour) is satisfied.  The linear
 section condition serves to reduce from the ten extended coordinates
 to four physical coordinates.

\subsection\NNisFive{$n=5$}The duality group is $G= Spin(5,5)$ and 
the coordinates form an $R_1={\bf16}$.  The section condition is
 $$
  \partial_{\a} \gamma^{a \a \b}  \partial_{\b} = 0 \komma  \quad a =
  1\dots 10, \ \a = 1 \dots 16 \punkt 
 \eqn
$$

 We now introduce a linear section condition 
 $$
 0 = \Lambda^\a(\gamma^{ab})_\a{}^\b \partial_\b = 0 \komma
 \eqn
$$
 which transforms in $\bar R_4={\bf45}$, involving a spinor $\L^\a$ in
 $\bar R_3={\bf16}$. If we further make the restriction
 that $\Lambda$ is pure we see that this constraint implies that 11
 components of  $\partial_\a$ are constrained to vanish leaving
 correctly 5 physical coordinates.  To see this consider  decomposing
 $\Lambda = \phi + u_{[ab]}  + t_{[ab cd]}$ and considering the
 special choice for which $\phi \neq 0$ is the only
 non-vanishing. Then in terms of the Mukai pairing we need   
$$
\eqalign{
 &(\Lambda, \omega_1 \wedge \omega_2 \partial ) = 0 \komma \cr
 & (\Lambda, \iota_{X_1}\iota_{X_2} \partial ) = 0 \komma \cr
 & (\Lambda, \iota_{X_1} (\omega_1 \wedge  \partial)  - \omega_1
 \wedge ( \iota_{X_1} \partial)  ) =0 \punkt \cr
}
\eqn
$$
 With this choice for $\Lambda$ we see that first of these  sets the
 3-form of $\partial$ to zero, the second is trivial and the final
 one sets the five-form to zero.  What remains unconstrained are the
 five coordinates in the direction  of the one-form. This is a
 solution to the section condition.     

\Yboxdim4pt
\Yvcentermath1

\subsection\NNisSix{$n=6$}The duality group is the split form of
 $E_{6}$ and the coordinates are in the ${\bf 27}$\foot\star{It
 seems that the number of point-like charge in five dimensions gives a
 counting of ${\bf 27} \oplus {\bf1}$ and provides an extra singlet which
 is unneeded or irrelevant. Probably it can be accommodated but is
 rather trivially projected out by the section condition.}.
The section conditions must eliminate 21 of these components.    

To match with the field content we decompose according to $SL(6)
\times SL(2)$ so that the derivatives decompose as 
$\overline{\bf 27} \rightarrow
(\overline{\bf15},{\bf1}) \oplus ({\bf6}, {\bf2})$:  
$$
\partial = \partial^{ab} + \partial_a^\a\punkt
\eqn
$$
Then the section conditions are given by a ${\bf27} \rightarrow
({\bf15},{\bf1}) \oplus (\overline{\bf6}, {\bf2} ) $: 
$$
 \e_{\a \b} \partial^{\a}_a \partial^\b_b  +
 \e_{abcdef} \partial^{cd} \partial^{ef}  = 0 \komma
 \quad  \partial^{ab}\partial_b^\b= 0\komma  \quad a = 1\dots 6\komma\quad \a
 = 1 \dots 2 \komma 
\eqn
 $$
In other words we are projecting the $(\overline{\bf 27}  \otimes
\overline{\bf 27})|_{\bf27}$ to zero.   

Now to build the linear section condition we start with a $\L$ in
$\bar R_3={\bf78}$ and impose the ``purity constraint'' 
$\L^2|_{\bf650} = 0 $.  Then for this restricted
choice we tensor with a derivative and demand $\L\*
|_{\bar R_4={\bf351'}} = 0$.  
We decompose the ${\bf78}$ according to the $SL(6) \times SL(2)$,
${\bf78}\rightarrow
({\bf1},{\bf3})\oplus({\bf35},{\bf1})\oplus({\bf20},{\bf2})$,
$$
\L \rightarrow \phi^{(\a \b)} + u_a{}^b  + w_{[abc]}{}^\a  \punkt 
\eqn
$$ 
The constraint decomposes as
${\bf650}\rightarrow({\bf1},{\bf1})\oplus({\bf70},{\bf2})
\oplus(\overline{\bf70},{\bf2})\oplus({\bf20},{\bf2})\oplus({\bf189},{\bf1})
\oplus({\bf35},{\bf1})\oplus({\bf35},{\bf3})$, where some non-trivial
$SL(6)$ representations are
$$
{\bf20}:\,\yng(1,1,1)\quad
{\bf21}:\,\yng(2)\quad
{\bf70}:\,\yng(2,1)\quad
{\bf84}:\,\yng(2,1,1,1)\quad
{\bf105}:\,\yng(2,1,1)\quad
{\bf189}:\,\yng(2,2,1,1)\quad
\eqn
$$
A representative of the solution of the constraint on $\L$ can be
taken as $u_a{}^b = w_{[abc]}{}^\a= 0$, $\phi^{12}=\phi^{22} = 0$,
$\phi^{11} \neq 0$. 

Now we consider this particular solution in the section equation in
the representation $\bar R_4={\bf351'}$, which decomposes as
${\bf351'}\rightarrow(\overline{\bf21},{\bf1})
\oplus(\overline{\bf15},{\bf3})\oplus(\overline{\bf84},{\bf2})
\oplus({\bf6},{\bf2})\oplus({\bf105},{\bf1})$.
The relevant $SL(6)\times SL(2)$ representations of the linear section
condition (considering $u=w=0$) are only
$(\overline{\bf15},{\bf3})\oplus({\bf6},{\bf2})$. This implies
$\*^{ab}=0$ and $\*^2_a=0$. Six directions remain.

We can check that under dimensional reduction this reduces to the pure
spinor constraint and associated linear section condition. This
entails doing a branching into $SO(5,5)$ and essentially keeping only
the ${\bf10_{-2}} \subset {\bf27}$ and the ${\bf16_{-3}}\subset
{\bf78}$. From the purity
constraint, $({\bf 78}  \otimes_S {\bf 78}) |_{\bf650}  = 0$, one
recovers  $({\bf 16_{-3}}  \otimes{\bf 16_{-3}}) |_{\bf10_{-6}} $
and then from the section condition    $({\bf27}\otimes {\bf78}) |_{\bf351}
= 0$ one indeed recovers $({\bf10_{-2}}\otimes {\bf16_{-3}})
|_{\overline{\bf{16}}_{\bf-5} } = 0$.

\subsection\SSnisSeven{$n=7$}For $E_7$, the ``pure spinor'' 
should be in $\bar R_3={\bf912}$. Start by
decomposing various modules in $SL(8)$ modules:
$$
\eqalign{
{\bf56}&\longrightarrow{\bf28}\oplus\overline{\bf28}\cr
{\bf133}&\longrightarrow{\bf63}\oplus{\bf70}\cr
{\bf912}&\longrightarrow{\bf36}\oplus\overline{\bf36}
                   \oplus{\bf420}\oplus\overline{\bf420}\cr
{\bf1463}&\longrightarrow{\bf1}\oplus{\bf336}\oplus\overline{\bf336}
                   \oplus{\bf720}\oplus{\bf70}\cr
{\bf8645}&\longrightarrow{\bf63}\oplus{\bf378}
           \oplus\overline{\bf378}\oplus{\bf2352}
           \oplus{\bf945}\oplus\overline{\bf945}\oplus{\bf3584}\cr
}
\eqn
$$
The Young tableaux for some non-obvious representations are:
$$
{\bf70}:\,\yng(1,1,1,1)\quad
{\bf420}:\,\yng(2,1,1,1,1)\quad
{\bf336}:\,\yng(2,2)\quad
{\bf720}:\,\yng(2,2,1,1,1,1)\quad
{\bf378}:\,\yng(2,1,1)\quad
{\bf2352}:\,\yng(2,2,2,1,1)\quad
{\bf945}:\,\yng(3,1,1,1,1,1)\quad
{\bf3584}:\,\yng(3,2,2,2,1,1,1)
\eqn
$$
The section condition as a bilinear condition on the derivatives reads
$\*_{ac}\*^{bc}-\fr8\d_a^b\*_{cd}\*^{cd}=0$,
$\*_{[ab}\*_{cd]}+\fr{24}\e_{abcdefgh}\*^{ef}\*^{gh}=0$. A representative of the
solution can be taken as the linear subspace spanned by $\*_{a8}$,
which breaks to $SL(7)$. Consider an object $\Lambda$ in ${\bf912}$,
constrained by $\Lambda^2|_{\bf1463}=0$. One solution is that
$\Lambda$ only sits as an $SL(7)$ singlet $\lambda_{88}$ in the
$\bf36$. 
Consider now a linear condition $\Lambda\*|_{{\bf8645}\oplus{\bf133}}=0$. 
With $\Lambda$ in $\bf36$
as above, the constraints are $\Lambda_{ab}\*^{cd}=0$ (${\bf945}\oplus{\bf63}$),
$\Lambda_{a[b}\*_{cd]}=0$ ($\bf378$). This is solved by $\*_{ab}$ as above.

\section\Conclusions{Conclusions}In this paper, we have studied a
couple of different, but connected, aspects of generalised
diffeomorphisms in the U-duality (exceptional) framework. We have
examined their algebraic structure and reducibility in a U-duality
covariant formalism, and demonstrated how to understand and formulate
the section condition in a linear way.

One of the most striking observations here is the appearance of the
representations forming tensor hierarchies or Borcherds algebras,
connected to form fields of different degrees in the dimensionally
reduced theory, as representations describing the infinite
reducibility of the generalised diffeomorphisms. 
The representation contents of these structures have
varied slightly between different authors, but we believe that our
predictions, that are algebraically unique, will provide the generic
structure. This question certainly merits further attention.

Although we have not been able to give a consistent algebra of
generalised diffeomorphisms based on $E_8$, it is striking that the
algebra is as close to working as it is. It is also remarkable that
the natural extrapolation of the reducibility of gauge parameters
produces the correct counting, including the dual
diffeomorphisms. Even if $E_8$ in itself is a complicated algebra, it
is still finite-dimensional, and
may provide a relatively simple means of studying dual gravity without
the introduction of infinite-dimensional algebras --- the caveat of
course being that the structure must be modified in some way to make
sense algebraically.

We think that the covariant treatment in the present paper opens a
route to a
classification of generalised geometries. 
Another urgent question is the extension to supergeometries. In order
to take that step, a tensor calculus with a spin connection has to be
invented. Such a formulation is at present unknown, but will be needed
since fermions transform under the compact subgroup $K(E_{n(n)})$
[\DamourKleinschmidtNicolai]. 

\acknowledgements Much of this work was done during the programme
``Mathematics and Applications of Branes in String and M-theory'' at
the Isaac Newton Institute, Cambridge. MC would like to thank
Jakob Palmkvist for discussions. 
DSB has benefited from discussions with Malcolm Perry, Hadi and Mahdi
Godazgar, Chris Hull, Charles Strickland-Constable, Dan Waldram and
Peter West. He is partially supported by an STFC rolling grant
ST/J000469/1. 
DCT is supported by an FWO-Vlaanderen
postdoctoral fellowship and 
this work is supported in part by the Belgian Federal Science Policy
Office through the Interuniversity Attraction Pole IAP VI/11 and by
FWO Vlaanderen through project G011410N.

\refout
\end

%% file: ruled.tex
\catcode`@=11                                   
\catcode`\|=12                                  
\catcode`\&=4                                   

\newcount\ncols         \ncols=\z@              
\newcount\nrows         \nrows=\z@              
\newcount\curcol        \curcol=\z@             
     
\newdimen\thinsize      \thinsize=0.6pt         
\newdimen\thicksize     \thicksize=1.5pt        

\newif\iftableinfo      \tableinfotrue          
\newif\ifcentertables   \centertablestrue       
%
%
     
\let\plaincr=\cr                        
\let\plainspan=\span                    
\let\plaintab=&                         
\let\lparen=(                           
\let\NX=\noexpand                       

     
\def\ruledtable{\relax                          
    \@BeginRuledTable                           
    \@RuledTable}


\def\@BeginRuledTable{
   \ncols=0\nrows=0                             
   \begingroup                                  
    \offinterlineskip                           
    \def~{\phantom{0}}
    \def\span{\plainspan\omit\relax\colcount\plainspan}
    \let\cr=\crrule                             
    \let\CR=\crthick                            
    \let\nr=\crnorule                           
    \let\|=\Vb                                  
%
%
    \ifx\tablestrut\undefined\relax             
    \else\let\tstrut=\tablestrut\fi             
    \catcode`\|=13 \catcode`\&=13\relax         
    \TableActive                                
    \curcol=1                                   
%
%
    \ifdim\tablewidth>-\maxdimen\relax          %
      \edef\@Halign{\NX\halign to \NX\tablewidth\NX\bgroup\TablePreamble}%
      \tabskip=0pt plus 1fil                    
    \else                                       %
      \edef\@Halign{\NX\halign\NX\bgroup\TablePreamble}%
      \tabskip=0pt                              
    \fi                                         %
%
%
    \ifcentertables                             
       \ifhmode\vskip 0pt\fi                    
       \line\bgroup\hss                         
    \else\hbox\bgroup                           
    \fi}


\long\def\@RuledTable#1\endruledtable{
   \vrule width\thicksize                       
     \vbox{\@Halign                             
       \thickrule                               
       #1\relax                                 
       \tstrut                                  
       \plaincr\thickrule                       
     \egroup}
   \vrule width\thicksize                       
   \ifcentertables\hss\fi\egroup                
  \endgroup                                     
  \global\tablewidth=-\maxdimen                 
  \iftableinfo                                  
      \immediate\write16{[Nrows=\the\nrows, Ncols=\the\ncols]}%
   \fi}
     

\def\TablePreamble{
   \linecount                           
   \TableItem{####}
   \plaintab\plaintab                   
   \TableItem{####}
   \plaincr}


\def\@TableItem#1{
   \hfil\tablespace                             
   #1\relax                                     
   \tablespace\hfil                             
    }%

\def\@tableright#1{
   \hfil\tablespace\relax               
   #1\relax                             
   \tablespace\relax}

\def\@tableleft#1{
   \tablespace\relax                    
   #1\relax                             
   \tablespace\hfil}

\let\TableItem=\@TableItem              
     
\def\RightJustifyTables{\let\TableItem=\@tableright}
\def\LeftJustifyTables{\let\TableItem=\@tableleft}
\def\NoJustifyTables{\let\TableItem=\@TableItem}

\def\LooseTables{\let\tablespace=\quad}
\def\TightTables{\let\tablespace=\space}
\LooseTables                                    

%

\newdimen\tablewidth    \tablewidth=-\maxdimen  


\def\setRuledStrut{
   \dimen@=\baselineskip                        
   \advance\dimen@ by-\normalbaselineskip       
   \ifdim\dimen@<.5ex \dimen@=.5ex\fi           
   \setbox0=\hbox{\lparen}
   \dimen1=\dimen@ \advance\dimen1 by \ht0      
   \dimen2=\dimen@ \advance\dimen2 by \dp0      
   \def\tstrut{\vrule height\dimen1 depth\dimen2 width\z@}%
   }%

\def\tstrut{\vrule height 3.1ex depth 1.2ex width 0pt}


\def\bigitem#1{
   \setbox0=\hbox{#1}
   \dimen1 =\ht0 \dimen2 =\dp0                  
   \dimen@ =\baselines@ve                       
   \advance\dimen@ by-\normalbaselineskip       
   \ifdim\dimen@<.25ex \dimen@=.25ex\fi         
   \advance\dimen1 by \dimen@                   
   \advance\dimen2 by \dimen@                   
   \vrule height\dimen1 depth\dimen2 width\z@   
   \copy0}

     
%

     
\def\nextcolumn#1{
   \plaintab\omit#1\relax\colcount              
   \plaintab}
     
\def\tab{
   \nextcolumn{\relax}}


\def\vb{
   \nextcolumn{\vrule width\thinsize}}

\def\Vb{
   \nextcolumn{\vrule width\thicksize}}


     
{\catcode`\|=13 \let|0
 \catcode`\&=13 \let&0
 \gdef\TableActive{\let|=\vb \let&=\tab}%
}


\def\crrule{\relax                      
   \tstrut                              
   \plaincr\tablerule                   
  }%

\def\crthick{\relax                     
   \tstrut                              
   \plaincr\thickrule                   
  }%
     
\def\crnorule{\relax                    
   \tstrut                              
   \plaincr                             
   }%
   

     
\def\tablerule{\noalign{\hrule height\thinsize depth 0pt}}%
\def\thickrule{\noalign{\hrule height\thicksize depth 0pt}}%


%
%
%
     

\def\linecount{\relax\global\ncols=\curcol      
   \global\curcol=1                             
   \global\advance\nrows by 1\relax}
     
\def\colcount{\relax                            %
   \global\advance\curcol by 1\relax}


\newdimen\parasize      \parasize=4in           

%

%

\def\begintable{\relax                          
    \@BeginRuledTable                           
    \@begintable}

\long\def\@begintable#1\endtable{
   \@RuledTable#1\endruledtable}


\catcode`@=12                                   


%% file: youngtab.tex
\catcode`\@11\relax
\newif\ify@autoscale \y@autoscaletrue \def\Yautoscale#1{\ifnum #1=0
  \y@autoscalefalse\else\y@autoscaletrue\fi}
\newdimen\y@b@xdim
\newdimen\y@boxdim \y@boxdim=13pt
\def\Yboxdim#1{\y@autoscalefalse\y@boxdim=#1}
\newdimen\y@linethick    \y@linethick=.3pt
\def\Ylinethick#1{\y@linethick=#1}
\newskip\y@interspace \y@interspace=0ex plus 0.3ex
\def\Yinterspace#1{\y@interspace=#1}
\newif\ify@vcenter   \y@vcenterfalse
\def\Yvcentermath#1{\ifnum #1=0 \y@vcenterfalse\else\y@vcentertrue\fi}
\newif\ify@stdtext   \y@stdtextfalse
\def\Ystdtext#1{\ifnum #1=0 \y@stdtextfalse\else\y@stdtexttrue\fi}
\newif\ify@enable@skew   \y@enable@skewfalse
\expandafter\ifx\csname enableskew\endcsname\relax
 \y@enable@skewfalse \else \y@enable@skewtrue\fi
\def\y@vr{\vrule height0.8\y@b@xdim width\y@linethick depth 0.2\y@b@xdim}
\def\y@emptybox{\y@vr\hbox to \y@b@xdim{\hfil}}
\ify@enable@skew
 \def\y@abcbox#1{\if :#1\else
   \y@vr\hbox to \y@b@xdim{\hfil#1\hfil}\fi}
 \def\y@mathabcbox#1{\if :#1\else
   \y@vr\hbox to \y@b@xdim{\hfil$#1$\hfil}\fi}
\else
 \def\y@abcbox#1{\y@vr\hbox to \y@b@xdim{\hfil#1\hfil}}
 \def\y@mathabcbox#1{\y@vr\hbox to \y@b@xdim{\hfil$#1$\hfil}}
\fi
\def\y@setdim{%
  \ify@autoscale%
   \ifvoid1\else\typeout{Package youngtab: box1 not free! Expect an
     error!}\fi%
   \setbox1=\hbox{A}\y@b@xdim=1.6\ht1 \setbox1=\hbox{}\box1%
  \else\y@b@xdim=\y@boxdim \advance\y@b@xdim by -2\y@linethick
  \fi}
\newcount\y@counter
\newif\ify@islastarg
\def\y@lastargtest#1,#2 {\if\space #2 \y@islastargtrue
  \else\y@islastargfalse\fi}
\def\y@emptyboxes#1{\y@counter=#1\loop\ifnum\y@counter>0
  \advance\y@counter by -1 \y@emptybox\repeat}
\def\y@nelineemptyboxes#1{%
  \vbox{%
    \hrule height\y@linethick%
    \hbox{\y@emptyboxes{#1}\y@vr}
    \hrule height\y@linethick}\vskip-\y@linethick}
\def\yng(#1){%
  \y@setdim%
  \hskip\y@interspace%
  \ifmmode\ify@vcenter\vcenter\fi\fi{%
  \y@lastargtest#1,
  \vbox{\offinterlineskip
    \ify@islastarg
     \y@nelineemptyboxes{#1}
    \else
     \y@ungempty(#1)
    \fi}}\hskip\y@interspace}
\def\y@ungempty(#1,#2){%
  \y@nelineemptyboxes{#1}
  \y@lastargtest#2,
  \ify@islastarg
   \y@nelineemptyboxes{#2}
  \else
   \y@ungempty(#2)
  \fi}
\def\y@nelettertest#1#2. {\if\space #2 \y@islastargtrue
  \else\y@islastargfalse\fi}
\def\y@abcboxes#1#2.{%
  \ify@stdtext\y@abcbox#1\else\y@mathabcbox#1\fi%
  \y@nelettertest #2.
  \ify@islastarg\unskip%
   \ify@stdtext\y@abcbox{#2}\else\y@mathabcbox{#2}\fi%
  \else\y@abcboxes#2.\fi}
 \newdimen\y@full@b@xdim
 \newcount\y@m@veright@cnt
\ify@enable@skew
 \def\y@get@m@veright@cnt#1#2.{%
   \if :#1 \advance\y@m@veright@cnt by 1\y@get@m@veright@cnt#2.\fi}
 \let\y@setdim@=\y@setdim
 \def\y@setdim{%
   \y@setdim@ \y@full@b@xdim=\y@b@xdim
   \advance\y@full@b@xdim by 1\y@linethick}
 \def\y@m@veright@ifskew#1{
   \y@m@veright@cnt=0 \y@get@m@veright@cnt#1.
   \moveright \y@m@veright@cnt\y@full@b@xdim}
\else
 \def\y@m@veright@ifskew#1{}
\fi
\def\y@nelineabcboxes#1{%
  \y@nelettertest #1.
  \ify@islastarg
   \y@m@veright@ifskew{#1}
    \vbox{
      \hrule height\y@linethick%
      \hbox{\ify@stdtext\y@abcbox#1\else\y@mathabcbox#1\fi\y@vr}
      \hrule height\y@linethick}\vskip-\y@linethick
  \else
   \y@m@veright@ifskew{#1}
    \vbox{
      \hrule height\y@linethick%
      \hbox{\y@abcboxes #1.\y@vr}%
      \hrule height\y@linethick}\vskip-\y@linethick
  \fi}
\def\young(#1){%
  \y@setdim%
  \hskip\y@interspace%
  \y@lastargtest#1,
  \ifmmode\ify@vcenter\vcenter\fi\fi{%
  \vbox{\offinterlineskip
    \ify@islastarg\y@nelineabcboxes{#1}%
    \else\y@ungabc(#1)%
    \fi}}\hskip\y@interspace}
\def\y@ungabc(#1,#2){%
  \y@nelineabcboxes{#1}%
  \y@lastargtest#2,
  \ify@islastarg\y@nelineabcboxes{#2}%
  \else\y@ungabc(#2)%
  \fi}
\catcode`\@12\relax
 